\documentclass[10pt,journal]{IEEEtran}
\usepackage{amssymb}
\usepackage{amsmath}
\usepackage{cite}
\usepackage{hyperref}
\usepackage{url}
\usepackage{xcolor}
\usepackage{cite,graphicx,amsmath,amssymb}
\usepackage{fancyhdr}
\usepackage{mdwmath}
\usepackage{mdwtab}
\usepackage{caption}
\usepackage{amsthm}
\usepackage{setspace}
\usepackage{algorithm}
\usepackage{algorithmic}
\usepackage{subfig}
\usepackage{multirow}
\usepackage{makecell}
\usepackage{mathtools}

\graphicspath{{./src/img/}}

\newtheorem{theorem}{Theorem}

\newtheorem{lemma}{Lemma}

\newtheorem{corollary}{Corollary}

\allowdisplaybreaks
\title{Sense-Then-Train: An Active-Sensing-Based Beam Training Design for Near-Field MIMO Systems}
\author{Hao~Jiang,~\IEEEmembership{Graduate Student Member,~IEEE,}
        Zhaolin~Wang,~\IEEEmembership{Graduate Student Member,~IEEE,} \\
        and Yuanwei~Liu,~\IEEEmembership{Fellow,~IEEE}
        
\thanks{
Part of this work has been accepted for publication by the IEEE International Conference on Communications, Denver, USA, June 2024 \cite{jiang2024active}.

Hao Jiang and Zhaolin Wang are with the School of Electronic Engineering
and Computer Science, Queen Mary University of London, London E1 4NS, U.K. (e-mail: \{hao.jiang, zhaolin.wang\}@qmul.ac.uk).

Yuanwei Liu is with the Department of Electrical and Electronic Engineering, The University of Hong Kong, Hong Kong (e-mail: yuanwei@hku.hk).
}
\vspace{-1cm}
}

\begin{document}
\maketitle
\begin{abstract}
An active-sensing-based sense-then-train (STT) scheme is proposed for beam training in near-field multiple-input multiple-output (MIMO) systems.
Compared to conventional codebook-based schemes, the proposed STT scheme is capable of not only addressing the complex spherical-wave propagation but also effectively exploiting the additional degrees-of-freedoms (DoFs). 
The STT scheme is tailored for both single-beam and multi-beam cases. 
1)~For the single-beam case, the STT scheme first utilizes a sensing phase to estimate a low-dimensional representation of the near-field MIMO channel in the truncated wavenumber domain.
Then, in the subsequent training phase, the neural network modules at transceivers are updated online to align beams, utilizing sequentially received ping-pong pilots.
This approach can efficiently obtain the aligned beam pair without relying on predefined codebooks or training datasets.
2)~For the multi-beam case, based on the single-beam STT, a Gram-Schmidt method is further utilized to guarantee the orthogonality between beams in the training phase.
Numerical results unveil that 1)~the proposed STT scheme can significantly enhance the beam training performance in the near field compared to the conventional far-field codebook-based schemes, and 2)~the proposed STT scheme can perform fast and low-complexity beam training, while achieving a near-optimal performance without full channel state information in both cases.
\end{abstract}

\section{Introduction}
As a further step from the fifth generation (5G) technologies, the next-generation mobile networks are envisioned to accommodate more than 15 billion mobile broadband (MBB) subscribers and support more than 250 gigabits traffic for every mobile user per month \cite{jiang2021road, itu2015imt}.
Due to the low-frequency bands tend to be saturated, the high-frequency bands, such as millimeter wave (mmWave) and terahertz (THz) bands, are anticipated as critical enablers for the next-generation mobile networks by providing an enormous bandwidth with an order of tens up to a hundred gigahertz (GHz) \cite{Shafie2023Therahertz, wang2018millimeter}.  

Nevertheless, communications over such high frequencies inevitably suffer from atmospheric-induced attenuation, leading to considerable throughput degradation.
As a remedy, extremely large-scale antenna arrays (ELAAs) can build highly reliable communication links using narrow beams, thus compensating for the propagation loss \cite{jiang2023active}.
However, this property of ELAAs leads to a dilemma for the system design. 
On the one hand, due to the small coverage of the narrow beams, a slight misalignment between beams and user channels can lead to significant performance loss \cite{nitsche2015steering}, rendering accurate channel state information (CSI) critical for ELAA systems. 
On the other hand, the extremely large channel dimensions in ELAA systems can lead to unacceptable pilot overheads to obtain accurate CSI using conventional channel estimation methods.
To address this issue, beam training has been proposed as an initial access method to establish stable communication links without CSI in the 5G new radio (NR) \cite{qurratulain2023machine, heng2023grid, Giordani2019tutorial}. 

However, another critical issue needed to be considered for ELAAs is new electromagnetic (EM) channel features, due to the large physical sizes.
More specifically, the EM field radiating from antennas can be divided into a near-field region and a far-field counterpart.
In the far-field region, the physical size of antennas is negligible compared to the distance between transceivers.
Therefore, the spherical-wave propagation can be approximated by the planar-wave propagation, leading to array responses being determined solely by angles.
In contrast, in the near-field region, due to the adjacency of transceivers, the planar-wave assumption fails to capture the EM characteristics since the array responses are non-uniform for a given angle.  
The boundary between the near- and far-field regions can be characterized by Rayleigh distance $\frac{2D^2}{\lambda}$, where $D$ and $\lambda$ denote the antenna aperture and the carrier frequency, respectively \cite{liu2023near}.
Accordingly, with larger apertures and high-frequency communications, ELAAs can extend the Rayleigh distance to hundreds of meters \cite{wu2023two, liu2024nearfield}, resulting in a larger near-field region.
Hence, it is critical to consider the prominent near-field characteristics to harness the high beam gain brought by ELAAs.
Moreover, in contrast to the rank-$1$ far-field line-of-sight (LoS) channel, the near-field LoS channel exhibits more available DoFs, i.e., ranks, due to the spherical-wave propagation.
These extra DoFs can be potentially exploited to support higher spectral efficiency (SE).

\subsection{Prior Works}
\subsubsection{Far-Field Beam Training}
The codebook-based beam training method is popular for establishing a stable communication link in far-field systems. 
Particularly, for the multiple-input single-output (MISO) case, the authors of \cite{song2017common} and \cite{zhang2020beam} designed an angular-domain codebook consisting of codewords, i.e., beams, pointing to specified angular directions.
Hence, the transmitter can find the strongest beam by sweeping the whole codebook.
Although this method is simple and straightforward in the MISO case, it will cause large training overhead for the multiple-input multiple-output (MIMO) case, since a beam pair instead of a single beam needs to be found by exhaustively searching codebooks on both sides.
To reduce the searching overheads, the authors of \cite{heng2023grid} and \cite{wang2023improving} conceived a site-specific codebook design for MIMO systems, in which codebooks are tailored for the specific network topology at a given site.
Even though the site-specific adoption can reduce the codebook size by excluding the beams that are not frequently used, the codebook, however, cannot cover the whole angular space, which must be redesigned when the network topology is changed. 
To address this issue, the authors of \cite{xiao2026hierarchical, qi2020hierarchical, he2015suboptimal} proposed hierarchical codebooks, which allow beams to be searched in a coarse-to-fine manner.
Therefore, the searching overhead for traversing the whole angular space is reduced to a logarithmic order.
However, this method suffers from the error propagation issue, due to the bisection search operation. 
To further enhance the accuracy of beam training, the authors of \cite{li2019explore} proposed a two-stage search method, which can be proved to asymptotically outperform both the hierarchical and the exhaustive searching schemes.
The above methods are all codebook-based, meaning that the resolution and the training overheads are limited by the codebook design.
Hence, to eliminate the need for codebooks, the authors of \cite{jiang2023active} and \cite{sohrabi2022active} proposed an active-sensing method, in which the transceivers align their beams using ping-pong pilots.
Besides, the active-sensing method further exploited recurrent neural networks (RNNs) to capture the temporal correlations between pilots, thus accelerating the beam training process. 

\subsubsection{Near-Field Beam Training}
Compared to the far-field scenario, research on near-field beam training is still in its infancy and solely focuses on the MISO case.
In near-field systems, the spherical-wave propagation in the near-field regions brings a new distance dimension to beam patterns.
Therefore, even in the same angular direction, the phase distribution of signals varies with the distance, which is fundamentally different from far-field systems, according to \cite{las2006evaluating} and \cite{liu2023near}.
By taking into account the new distance dimension, the authors of \cite{cui2022channel} first proposed a polar-domain codebook by sampling the angular space uniformly and the distance space non-uniformly.
After this codebook is swept exhaustively, the strongest beam can be found by the transmitter.
To reduce the searching overheads caused by exhaustive sweeping, the authors of \cite{zhang2022fast} and \cite{wu2023two} proposed a two-stage beam training scheme, in which two-dimensional search was converted to two sequential phases.
Particularly, the angular space was first traversed using conventional far-field codebooks to estimate the coarse angle of the user.
Then, in the next phase, the customized polar-domain codebook was employed to find the distance of the user.
Since only a sub-set in the angular space is swept, searching overhead can be reduced.
As separate studies, \cite{wu2023two} and \cite{zhang2023beam} proposed hierarchical near-field codebook designs to reduce the searching overhead, in which the polar domain was searched via hierarchical codebook in an angular-then-distance manner. 
For wideband near-field networks, the authors of \cite{cui2023near-rainbow} proposed a near-field rainbow scheme to accelerate beam training by exploiting the beam split effect.
In this method, the angular domain was searched in a frequency-division manner instead of a time-division manner, thus speeding up beam training.

\subsection{Motivations and Contributions}
Although beam training for near-field MISO systems has been extensively investigated in the literature, beam training for near-field MIMO systems has not been studied to the best of the authors' knowledge. 
In contrast to MISO systems and far-field MIMO systems, the codebook-based method is no longer suitable for near-field MIMO systems. In particular, in both far-field and near-field MISO systems, the codewords can be designed by only considering the transmit array response \cite{zhang2022fast}. 
For far-field MIMO systems, both LoS and non-line-of-sight (NLoS) channel matrices can be decomposed into the independent transmit and receive array response vectors, thus facilitating the independent codebook design at the transmitter and receiver, respectively \cite{liu2023near, ouyang2023nearfield}. 
However, in near-field MIMO systems, such a decomposition is no longer valid for the LoS channel since the transmit and receive array responses are highly coupled due to the spherical-wave propagation.
Therefore, the codebook design for this case is challenging and non-trivial. 
As a remedy, the codebook-free active-sensing method based on neural networks (NNs) is promising to address the aforementioned challenges. 
Although the canonical active-sensing method proposed in \cite{jiang2023active} and \cite{sohrabi2022active} can facilitate efficient beam training, it suffers from three main drawbacks when applied to near-field MIMO systems.
1)~\textbf{High Computational Complexity:~}This method executes beam training directly on the space-domain channel representations, thus resulting in a computational complexity scaling with the number of antennas.
However, due to ELAAs' large number of antennas, this method is less tractable due to high computational complexity.
2)~\textbf{Unexploited DoFs: }This method can only find one single pair of beams corresponding to the largest singular value of channel matrices.
Thus, the additional DoFs offered by the near-field channel are unexploited.
3)~\textbf{Inflexibility in Dynamic Environment:} The offline training and online implementation framework used by the canonical active-sensing method is not applicable to the scenarios where the dimension of the outputs varies.
To address these issues, we propose a novel sense-then-train (STT) beam training scheme for near-field MIMO systems, which not only reduces the complexity via sensing but also can be applied to the multi-beam case to utilize the additional DoFs.
The contribution of this work is summarized as follows:
\begin{itemize}
		\item We propose a codebook-free STT beam training scheme for near-field MIMO systems.
		To circumvent the high computational complexity incurred by the high-dimension space-domain channel representations, the proposed method facilitates beam training in the low-dimensional subspace of the wavenumber domain.
		To this end, prior to the beam training phase, a sensing phase is employed to obtain the truncated wavenumber-domain transformation matrices (WTMs), whose dimensions are determined jointly by the DoFs of near-field channels and the sensing thresholds.
		Then, a pair of beam training methods is proposed to obtain the beam(s) in the truncated wavenumber domain for both single-beam and multi-beam cases.
		\item For the single-beam STT scheme, an active-sensing-based method is employed for beam training without CSI, thus eliminating the necessity of predefined codebooks.
		Since the dimensions of WTMs cannot be determined in advance, the neural networks are initialized according to WTMs obtained via sensing and then trained incrementally in an online fashion.
		\item For the multi-beam STT scheme, based on the single-beam STT scheme, a Gram-Schmidt method is further exploited to mitigate inter-beam interference by ensuring the orthogonality between beams.
		Therefore, with such a method, multiple beams can be trained in a successive manner, thus enabling an adequate use of the available DoFs in near-field regions.
		\item  Numerical results unveil that i) with the aid of sensing results, the proposed STT scheme can facilitate fast and low-complexity beam training in the wavenumber domain, and ii) the proposed STT scheme can achieve
		near-optimal performance in both the single-beam and multi-beam cases, compared to the benchmark algorithm that necessitates perfect CSI.
\end{itemize}

\subsection{Organization and Notations}
The remainder of this work is organized as follows.
Section \ref{sect:system model} presents the near-field beam training system model and briefly introduces the wavenumber-domain transformation.
Sections \ref{sect:SDS} and \ref{sect:MDS} elaborate on the single-beam and multi-beam STT schemes, respectively.
In Section \ref{sec:result}, numerical results are provided to verify the effectiveness of our method.
Lastly, the conclusions are drawn in Section \ref{sec:conclusion}.

\textit{Notations:}
Scalars, vectors, and matrices are denoted by the lower-case, bold-face lower-case, and bold-face upper-case letters, respectively.
$\mathbb{C}^{M \times N}$ and $\mathbb{R}^{M \times N}$ denote the space of $M \times N$ complex and real matrices, respectively.
$(\cdot)^T$, $(\cdot)^*$, $(\cdot)^H$, and $\mathrm{Tr}(\cdot)$ denote the transpose, conjugate, conjugate transpose, and trace, respectively.
$|\cdot|$ represent determinant or absolute value depending on context.
$(\cdot)^{\left| \cdot \right|}$ and $\oslash$ denote the element-wise magnitude and division, respectively.
For a matrix $\mathbf{A}$, $[\mathbf{A}]_{:,j}$, $[\mathbf{A}]_{i,:}$, and $[\mathbf{A}]_{i,j}$ denote the $j$-th column, the $i$-th row, and the $(i,j)$-th element, respectively.
For a vector $\mathbf{a}$, $[\mathbf{a}]_i$ and $\left\| \mathbf{a} \right\| $ denote the $i$-th element and $2$-norm, respectively.
For a scalar $a$, $\lfloor a \rfloor$ and $\lceil  a \rceil $ denote the flooring and ceiling functions, respectively.

\section{System Model} \label{sect:system model}
\begin{figure}[t!]
	\centering
	\includegraphics[width=0.4\textwidth]{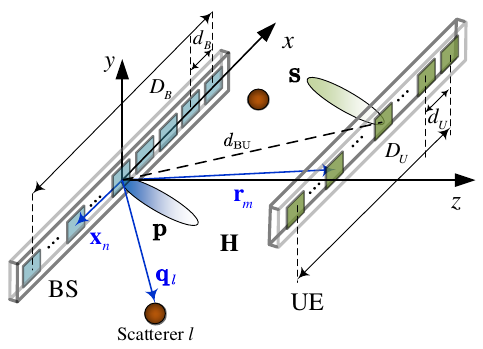}
	\caption{An illustration of a near-field MIMO system.}
	\label{fig:system_model}
\end{figure}
We consider a narrowband near-field MIMO communication system as depicted in Fig. \ref{fig:system_model}, which consists of a base station (BS) equipped with an $N$-antenna uniform linear array (ULA) with $N = 2 \tilde{N} - 1$ and an user equipment (UE) with an $M$-antenna ULA with $M = 2 \tilde{M} - 1$.
To enhance energy efficiency, hybrid analog and digital beamforming architecture is considered in this paper, in which only $N_{\mathrm{RF}} \ll N$ radio frequency (RF) chains are employed at the BS and the UE. 
Each RF chain is connected to all antennas through a high-dimensional analog beamforming network implemented by the low-cost phase shifters (PSs).  
Furthermore, the ULAs are assumed to be located on the $xz$ plane. 
The array apertures of the ULAs at the BS and the UE can be calculated by $D_{\mathrm{B}} = (N - 1)d_{\mathrm{B}}$ and $D_{\mathrm{U}} = (M-1)d_{\mathrm{U}}$, respectively, with $d_{\mathrm{B}}$ and $d_{\text{U}}$ denoting the antenna spacing.
The distance between the center points of the BS and the UE is assumed to be shorter than Rayleigh distance, i.e., $d_{\mathrm{BU}} < \frac{2(D_{\mathrm{B}} + D_{\mathrm{U}})^2}{\lambda}$, where $\lambda$ denotes the signal wavelength, but larger than the boundary of the reactive near-field region.

\subsection{Channel Representation in Space Domain}
The uniform spherical-wave (USW) channel model \cite{liu2023near} is adopted to model the near-field MIMO channel, which consists of a LoS link and $L$ NLoS links caused by randomly deployed scatterers.
Let ${\bf x}_n = [x_x^{(n)}, x_y^{(n)},  x_z^{(n)}]^T \in \mathbb{R}^{3 \times 1}$, where $n \in \{-\tilde{N}, ..., \tilde{N}\}$, denote the coordinate of the $n$-th antenna at the BS, ${\bf r}_m = [r_x^{(m)}, r_y^{(m)},  r_z^{(m)}]^T \in \mathbb{R}^{3 \times 1}$, where $ m \in\{-\tilde{M}, ..., \tilde{M}\}$ denotes the $m$-th antenna at the UE, and $\mathbf{q}_l=[ q_{x}^{(l)},q_{y}^{(l)},q_{z}^{(l)} ]^T \in \mathbb{R}^{3 \times 1}$ denotes the coordinate of the $l$-th scatterer.
It is noted that according to the USW model, the channel gains between the transmitter and receiver are approximated by that of the central link, which is denoted by $\beta$ \cite{wang2023near}.
In particular, let $\zeta _{\mathrm{pathloss}}\left( f,d \right) =( 4\pi fd/c )^2\exp \left(\varrho \left( f \right) d \right)$ denote the pathloss, where $\varrho \left( f \right)$ denotes the frequency-dependent medium absorption coefficient found in the dataset of \cite{ROTHMAN20134} and $d$ denotes the distance.
Therefore, the channel gain can be written as $\beta =\zeta _{\mathrm{pathloss}}^{-1}\left( f,d_{\mathrm{BU}} \right) G_{\mathrm{t}}G_{\mathrm{r}}$, where $G_{\mathrm{t}}$ and $G_{\mathrm{r}}$ denote the transmit and receive antenna gains, respectively.
Therefore, the $(m,n)$-th entry of the LoS channel matrix between the BS and the UE can be characterized by
\begin{align}
	\left[ \mathbf{H}_{\mathrm{LoS}} \right] _{m,n}=\beta e^{-jk_0\left\| \mathbf{r}_m-\mathbf{x}_n \right\|},
\end{align} 
where $k_0\triangleq 2\pi /\lambda$ denotes the wavenumber.
On the contrary, the NLoS components can be written as a multiplication of transmit and receive response vectors as follows:
\begin{align}
	\mathbf{H}_{\mathrm{NLoS}}=\sum_{l=1}^L{\beta _l\mathbf{b}_{\mathrm{B}}\left( \mathbf{q}_l \right) \mathbf{b}_{\mathrm{U}}^{T}\left( \mathbf{q}_l \right)},
\end{align}
where $\beta _l$ denotes the channel gain of the $l$-th NLoS channel.
In particular, $\beta _l$ can be calculated as $\beta _l=\alpha_l \zeta _{\mathrm{pathloss}}^{-1}\left( f,r_{l} \right) G_{\mathrm{t}}G_{\mathrm{r}}$, where $r_{l}$ and $\alpha_l$ denote the length and the scattering loss of the $l$-th NLoS link.
Vectors $\mathbf{b}_{\mathrm{B}}\left( \mathbf{q}_l \right) \in \mathbb{C}^{N \times 1}$ and $\mathbf{b}_{\mathrm{U}}\left( \mathbf{q}_l \right) \in \mathbb{C}^{M \times 1}$ denote the array response vectors at the BS and the UE, respectively, which can be expressed as follows:
\begin{align}
	\mathbf{b}_{\mathrm{B}}\left( \mathbf{q}_l \right) &=\left[ e^{-jk_0\left\| \mathbf{q}_l-\mathbf{x}_{-\tilde{N}} \right\|},...,e^{-jk_0\left\| \mathbf{q}_l-\mathbf{x}_{\tilde{N}} \right\|} \right] ^T, \\
	\mathbf{b}_{\mathrm{U}}\left( \mathbf{q}_l \right) &=\left[ e^{-jk_0\left\| \mathbf{q}_l-\mathbf{r}_{-\tilde{M}} \right\|},...,e^{-jk_0\left\| \mathbf{q}_l-\mathbf{r}_{\tilde{M}} \right\|} \right] ^T.
\end{align} 
By considering both of the above, the near-field MIMO channel can be written as
\begin{align}
	\mathbf{H}=\mathbf{H}_{\mathrm{LoS}} + \mathbf{H}_{\mathrm{NLoS}}.
\end{align}
It is noted that in high-frequency bands, communication channels are LoS-dominated and NLoS-assisted due to severe scattering losses \cite{tang2023line}.
Therefore, for the far-field scenario, the rank of the channel is approximately one, since it can be expressed as a production of transceivers' array responses.
On the contrary, for the near-field scenario, the rank of the channel is distance-dependent but larger than one. 
However, for a typical communication distance, the rank of the channel is still much smaller than the dimension of the channel, indicating that there is some redundant information. 

\subsection{Channel Representation in Wavenumber Domain}
To remove the redundant information in near-field MIMO channels, the wavenumber-domain representation can be exploited.
The basic idea of this transformation is that a spherical wavefront in the near-field region can be approximated by a superposition of multiple planar wavefronts.  
According to the methodology in \cite{pizzo2021fourier,tang2023line, wei2022multi}, the spatial impulse response between the $n$-th antenna at the BS and the $m$-th antenna at the UE, i.e., $h_{m,n} = \left[ \mathbf{H} \right] _{m,n}$, can be derived based on a four-dimensional (4D) Fourier plane-wave representation, which is given by
\begin{align}
	&h_{m,n}=\frac{1}{\left( 2\pi \right) ^2}\iiiint\limits_{\mathcal{D} _{\boldsymbol{\kappa }}\times \mathcal{D} _{\mathbf{k}}}^{}{a_{\mathrm{U}}\left( \boldsymbol{\kappa },\mathbf{r}_m \right) h_{\mathrm{a}}\left( \kappa _x,\kappa _y,k_x,k_y \right)} \notag \\
	&\qquad \qquad \qquad \qquad \qquad ~ \times a_{\mathrm{B}}\left( \mathbf{k},\mathbf{x}_n \right) d\kappa _xd\kappa _ydk_xdk_x. \label{eq:ft_cont}
\end{align}
Here, $h_{\mathrm{a}}\left( \kappa _x,\kappa _y,k_x,k_y \right) $ denotes the coupling coefficient between the transmit and receive plane waves, that are respectively given by $\mathbf{k}=[k_x,k_y,\gamma \left( k_x,k_y \right) ]^T$ and $\boldsymbol{\kappa }=[\kappa _x,\kappa _y,\gamma \left( \kappa _x,\kappa _y \right) ]^T$, in which $\gamma \left( a,b \right) \triangleq \sqrt{k_0^2-a^2-b^2}$.
Scalars $a_{\mathrm{B}}\left( \mathbf{k},\mathbf{x}_n \right) =e^{-j\mathbf{k }^T\mathbf{x}_n}$ and $a_{\mathrm{U}}\left( \boldsymbol{\kappa },\mathbf{r}_m \right) =e^{j\boldsymbol{\kappa }^T\mathbf{r}_m}$ denote the transmit and receive responses at the $n$-th antenna at the BS and the $m$-th antenna at the UE, respectively
In the radiating near-field region, $\gamma ( k_x,k_y ) $ and $\gamma ( \kappa_x,  \kappa_y  ) $ are real-valued.
Then, wavenumber domains at the BS and the UE can be respectively defined as
\begin{align}
	\mathcal{D} _{\mathbf{k}} &\triangleq \left\{ ( k_x,k_y ) \in \mathbb{R} ^2: k_x ^2+k_y^2\leqslant k_0^2 \right\}, \label{eq:D_k}
	\\
	\mathcal{D} _{\boldsymbol{\kappa }} &\triangleq \left\{ ( \kappa_x, \kappa_y ) \in \mathbb{R} ^2:  \kappa_x ^2+ \kappa_y^2\leqslant k_0^2 \right\}.  \label{eq:D_kap}
\end{align}
According to \cite[Theorem 2]{pizzo2021fourier}, when antenna arrays are electromagnetically large, MIMO channels can be approximated by a finite number of plane waves.
In this paper, since the ULAs are assumed to be deployed on the $xz$-plane, we have $r_{y}^{(m)}=x_{y}^{(n)}=0$.
In this case, the transmit and receive responses can be simplified by $a_{\mathrm{B}}( \mathbf{k},\mathbf{x}_n )=\exp({-j( k_xx_{x}^{( n )}+\gamma ( k_x ) x_{z}^{( n )})})
$ and $a_{\mathrm{U}}( \boldsymbol{\kappa },\mathbf{r}_m )=\exp({-j( \kappa _xr_{x}^{( m )}+\gamma ( \kappa _x ) r_{z}^{( m )} )})$, indicating that $k_y$ and $\kappa_y$ can be discarded.
Therefore, we have $-k_0 \le k_x, \kappa_x \le +k_0$.
Then by sampling $k_x$ and $\kappa_x$ with intervals of $2\pi / D_{\rm B}$ and $2\pi/ D_{\rm U}$, respectively,  
the discrete version of \eqref{eq:D_k} and \eqref{eq:D_kap} can be written as $\mathcal{G} _{\mathbf{k}}\triangleq \{j\in \mathbb{Z} :-k_0\leqslant 2 \pi j/D_{\mathrm{B}}\leqslant +k_0\}
$ and $\mathcal{G} _{\boldsymbol{\kappa }}\triangleq \{i\in \mathbb{Z} : -k_0\leqslant 2\pi i/D_{\mathrm{U}}\leqslant +k_0\}
$, respectively.
With these discrete sets, the spatial impulse response $h_{m,n}$ in \eqref{eq:ft_cont} can be discretized to   
\begin{align}
	&h_{m,n}  \approx \notag \\
	&\sum_{i\in \mathcal{G} _{\boldsymbol{\kappa }}}{\sum_{j\in \mathcal{G} _{\mathbf{k}}}{\phi _{\mathrm{U}}\left( i,r_{x}^{(m)} \right) \tilde{h}_{\mathrm{a}}\left( i,j,r_{z}^{(m)},x_{z}^{(n)} \right) \phi _{\mathrm{B}}^{*}\left( j,x_{x}^{(n)} \right)}}
	, \label{eq:H_discrete} 
\end{align}
where
\begin{align}
	\tilde{h}_{\mathrm{a}}\left( i,j,r_{z}^{(m)},x_{z}^{(n)} \right) &=e^{j\gamma \left( \frac{2\pi i}{D_{\mathrm{U}}} \right) r_{z}^{(m)}}h_{\mathrm{a}}\left( i,j \right) e^{-j\gamma \left( \frac{2\pi j}{D_{\mathrm{B}}} \right) x_{z}^{(n)}}
	, \label{eq:H_discrete-1}\\
	\phi _{\mathrm{U}}\left( i,r_{x}^{(m)} \right) &=\exp \left( {j\frac{2\pi i}{D_{\mathrm{U}}}r_{x}^{(m)}} \right),
	\label{eq:H_discrete-2} \\
	\phi _{\mathrm{B}}\left( j,x_{x}^{(n)} \right) &=\exp \left( {j\frac{2\pi j}{D_{\mathrm{B}}}x_{x}^{(n)}} \right). \label{eq:H_discrete-3}
\end{align} 
Based on the above transformation, the overall near-field MIMO channel can be approximated by 
\begin{align}
	\mathbf{H} \approx \sqrt{MN} \mathbf{\Phi }_{\mathrm{U}}\tilde{\mathbf{H}}_{\mathrm{a}}\mathbf{\Phi }_{\mathrm{B}}^{H} \label{eq:H_a}.
\end{align}
In the expression, $[ \tilde{\mathbf{H}}_{\mathrm{a}} ] _{i,j}=\tilde{h}_{\mathrm{a}}(i,j,r_{z}^{(m)},x_{z}^{(n)})$, $\mathbf{\Phi }_{\mathrm{U}}$ and $\mathbf{\Phi }_{\mathrm{B}}$ are the semi-unitary wavenumber-domain transform matrices (WTMs) given by
\begin{align}
	\mathbf{\Phi }_{\mathrm{U}}&=\left[ \boldsymbol{\phi }_{\mathrm{U},\lceil -D_{\mathrm{U}}/\lambda \rceil}^{},...,\boldsymbol{\phi }_{\mathrm{U},\lfloor +D_{\mathrm{U}}/\lambda \rfloor}^{} \right] \in {\mathbb{C} ^{M\times \left| \mathcal{G} _{\boldsymbol{\kappa }} \right|}}, \label{eq:WTM_Rx}
	\\
	\mathbf{\Phi }_{\mathrm{B}}&=\left[ \boldsymbol{\phi }_{\mathrm{B},\lceil -D_{\mathrm{B}}/\lambda \rceil}^{},...,\boldsymbol{\phi }_{\mathrm{B},\lfloor +D_{\mathrm{B}}/\lambda \rfloor}^{} \right] \in \mathbb{C} ^{N\times \left| \mathcal{G} _{\mathbf{k}} \right|} , \label{eq:WTM_Tx}
\end{align}
where $\boldsymbol{\phi }_{\mathrm{U},i}^{}=\frac{1}{\sqrt{M}}[ \phi _{\mathrm{U}}^{}( i,r_{x}^{(1)} ) ,...,\phi _{\mathrm{U}}^{}( i,r_{x}^{(M)} ) ] ^T\in \mathbb{C} ^{M\times 1}
$ and $\boldsymbol{\phi }_{\mathrm{B},j}^{}=\frac{1}{\sqrt{N}}[ \phi _{\mathrm{B}}^{}( j,x_{x}^{(1)}) ,...,\phi _{\mathrm{B}}^{}( j,x_{x}^{(N)} )] ^T\in \mathbb{C} ^{N\times 1}$.
According to \cite{pizzo2021fourier}, $\tilde{\mathbf{H}}_{\mathrm{a}} $  is semi-unitary equivalent to ${\mathbf{H}}$, meaning that they have the identical top singular values.
Moreover, compared with ${\mathbf{H}}$, $\tilde{\mathbf{H}}_{\mathrm{a}}$ has a diagonal and sparse structure.
The diagonal elements of $\tilde{\mathbf{H}}_{\mathrm{a}}$ represent the coupling coefficients between the planar wavefront at the transceivers.
By using the definition of semi-unitary, we have $\tilde{\mathbf{H}}_{\mathrm{a}}\approx \frac{1}{\sqrt{MN}}\mathbf{\Phi }_{\mathrm{U}}^{H}\mathbf{H\Phi }_{\mathrm{B}}^{}$.

\begin{figure}[t]
	\centering
	\includegraphics[width=0.7\linewidth]{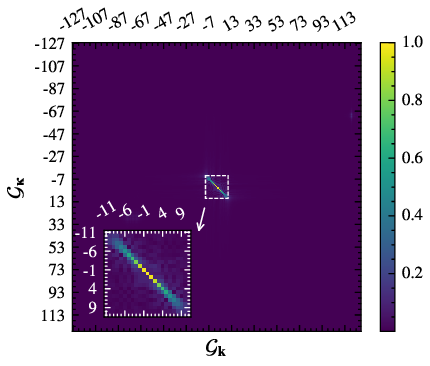}
	\caption{An illustration of channel representation in the wavenumber domain, i.e., $|\tilde{\mathbf{H}}_{\mathrm{a}}|$, under $M=N=255$, $d_{\rm BU}=15~m$, and $f=28~{\rm GHz}$. }
	\label{fig:H_a} 
\end{figure}
In Fig. \ref{fig:H_a}, we illustrate the normalized magnitude of $\tilde{\mathbf{H}}_{\mathrm{a}}$.
It can be observed that $\tilde{\mathbf{H}}_{\mathrm{a}}$ is sparse and diagonal, indicating that the redundant information in space-domain channel representations can be removed in the wavenumber domain.
Moreover, the significant values representing the dominant LoS components are located in a low-dimensional sub-space at the center of $\tilde{\mathbf{H}}_{\mathrm{a}}$. 
It is noted that the far-field channel is a special case of a near-field channel and is dominated by one planar wavefront. 
Consequently, the wavenumber domain analysis is applicable to the far-field scenarios.
The key distinction lies in that only one spatial DoF is available due to the planar-wave propagation in the far field.

\subsection{Problem Formulation}
We assume that there are $N_{\rm s}$ data streams transmitted from the BS to the UE through the MIMO channel. To effectively support these data streams with the minimum cost, we set $N_{\mathrm{RF}} = N_{\rm s}$. Let  $\mathbf{c} \in \mathbb{C}^{N_s \times 1}$, $\mathbf{\Lambda } \in \mathbb{R}^{N_{\rm s} \times N_{\rm s}}$, $\mathbf{P} \in {\mathbb{C}}^{N \times N_{\rm s}}$, and $\mathbf{S} \in {\mathbb{C}}^{M \times N_{\rm s}}$ denote the unit-power transmit signal, digital beamformer at the BS, analog beamformer at the BS, and analog combiner at the UE, respectively. 
In particular, the information symbols for each data stream are assumed to be independent and identically distributed, i.e., $\mathbb{E}[\mathbf{c}\mathbf{c}^H] = \mathbf{I}_{N_s}$. 
To facilitate beam training design, the digital beamformer is designed to only consider the power allocation between different beams, resulting in a diagonal structure of $\mathbf{\Lambda }$. 
Then, the received signal at the UE can be modeled as follows:
\begin{align}
	\mathbf{y}=\mathbf{S}^H\mathbf{HP}{\mathbf{\Lambda}} {\mathbf{c}}+\mathbf{S}^H\mathbf{n},
\end{align}  
where $\mathbf{n} \sim \mathcal{CN}(\boldsymbol{0}, \sigma^2 \mathbf{I}_{N_{\rm s}})$ denotes the complex Gaussian noise.
The SE of the considered near-field MIMO system is thus given by
\begin{align}
	R(\mathbf{S},\mathbf{P},\mathbf{\Lambda })=\log \left| \mathbf{I}_{N_{\mathrm{s}}}+\mathbf{C}^{-1}\mathbf{S}_{}^{H}\mathbf{HP}_{}\mathbf{\Lambda }\mathbf{\Lambda }^{H}\mathbf{P}_{}^{H}\mathbf{H}^H\mathbf{S}_{} \right|, \label{eq:se}
\end{align}
where $\mathbf{C}= \sigma ^2 \mathbf{S}^H\mathbf{S}$.
Therefore, the beam training problem can be formulated as:
\begin{subequations}
	\begin{align} 
		&\max_{\mathbf{S},\mathbf{P}, \mathbf{\Lambda }} R(\mathbf{S},\mathbf{P},\mathbf{\Lambda })\,\, 
		   \label{eq:P}\\ 
		&~{\rm{s.t.}}~ \left| [ \mathbf{P} ] _{i,j} \right|=\frac{1}{\sqrt{N}}, ~~\forall i,j, \label{eq:P-c1}\\
		&~\phantom{{\rm{s.t.}}}~\left| [ \mathbf{S} ] _{i,j} \right|=\frac{1}{\sqrt{M}}, ~~\forall i,j,  \textbf{\label{eq:P-c2}} \\
		&~\phantom{{\rm{s.t.}}}~\mathrm{Tr}\left\{ {\mathbf{\Lambda}}{\mathbf{\Lambda}}^{H} \right\} =P_{\mathrm{B}},
	\end{align}
\end{subequations}
where $P_{\rm B}$ denotes the transmit power at the BS.
To solve this problem, we proposed an active-sensing-based method to obtain the beamformers without needing CSI or codebooks.
It is important to underscore that the neural networks in the proposed method are established according to the dimension of truncated WTMs and trained in an online manner. 
Then, the power allocation matrix is designed based on the obtained analog beamformers. 
In the following, we first investigate problem \eqref{eq:P} for the simplest single-beam case, i.e., $N_{\rm s} = 1$, in Section \ref{sect:SDS}. Then, the general multi-beam case is studied in Section \ref{sect:MDS}.

\section{STT for Single-Beam Cases} \label{sect:SDS}
For the single-beam case, i.e., $N_{\rm s} = 1$, the SE in \eqref{eq:se} reduces to 
\begin{align}
	R(\mathbf{s},\mathbf{p})=\log \left( 1+\frac{\left| \mathbf{s}^H\mathbf{Hp} \right|^2}{\sigma ^2} \right).
\end{align}
Accordingly, problem \eqref{eq:P} can be rewritten as:
\begin{subequations}
	\begin{align} 
		&\max_{\mathbf{s}, \mathbf{p}}~\left| \mathbf{s}^H\mathbf{Hp} \right|^2 \label{eq:SD-Obj} \\ 
		&~{\rm{s.t.}}~ \eqref{eq:P-c1}{\rm~and~}\eqref{eq:P-c2}. \notag
	\end{align}
\end{subequations}
To obtain the near-optimal $\mathbf{s}$ and $\mathbf{p}$ of problem \eqref{eq:SD-Obj}, a proposed STT beam training scheme is first carried out before data transmission. 
In the following, the transmission protocol and detailed implementation of the proposed STT scheme will be provided.

\subsection{Signal Model and Transmission Protocol}
The STT scheme utilizes a ping-pong pilot scheme, where the transceivers transmit pilots alternatively.
Specifically, the BS first transmits a pilot to the UE via downlink (DL).
Then, after this pilot is received, the UE will respond by transmitting an alternative pilot to the BS via uplink (UL).
The pair of pilots is called ping-pong pilots, and such a round is called a ping-pong round, which is indexed by $t$.

Let $\mathbf{p}_t \in \mathbb{C}^{N \times 1}$ and $\mathbf{s}_t \in \mathbb{C}^{M \times 1}$ denote the analog transmit and receive beamformers at the BS and the UE at round $t$, respectively.
Assuming that the reciprocity of $\mathbf{H}$ holds, the received signal at the UE in DL and that at the BS in UL can be expressed as
\begin{align}
	{\mathbf{y}}_t^{\mathrm{DL}}&=\mathbf{H}{\mathbf{p}_t}c_{\mathrm{B}, t}+{\bf n}_{\mathrm{U}, t}, \\
	{\mathbf{y}}_t^{\mathrm{UL}}&=\mathbf{H}^T\mathbf{s}_t^* c_{\mathrm{U}, t}+{\bf n}_{\mathrm{B}, t},
\end{align}
where $c_{\mathrm{B}, t} \in \mathbb{C}$ and  $c_{\mathrm{U}, t} \in \mathbb{C}$ denote the baseband pilot signals at the BS and the UE, respectively, satisfying $|c_{\text{B}, t}|^2 = P_\text{B}$ and $|c_{\text{U}, t}|^2 = P_\text{U}$, and 
$\mathbf{n}_{\mathrm{B}} \sim \mathcal{CN}(0, \sigma^2 {\mathbf{I}}_N)$ and $\mathbf{n}_{\mathrm{U}} \sim \mathcal{CN}(0, \sigma^2{\mathbf{I}}_M )$ denote the complex Gaussian noise.  

\begin{figure}
	\centering
	\includegraphics[width=1\linewidth]{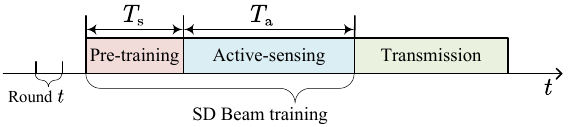}
	\caption{An illustration of the single-beam STT scheme.} \label{fig:STT_single-beam}
\end{figure}

The proposed STT scheme has two phases, namely, a sensing phase and a training phase. 
In the sensing phase, the truncated WTMs are estimated to reduce the training complexity.
Then, in the training phase, an active-sensing-based algorithm is proposed to obtain the optimal $\mathbf{s}$ and $\mathbf{p}$ based on the sensing results, i.e., truncated WTMs.
The duration of the former and the latter phases are denoted by $T_{\rm s}$ and $T_{\rm a}$, respectively.
In addition, the overall procedure of the single-beam STT scheme is illustrated by Fig. \ref{fig:STT_single-beam}, where ``SD" refers to ``single data stream". 

\subsection{Sensing Phase}\label{Sect:pre_training_SD}
As shown by Fig. \ref{fig:H_a}, the most significant channel powers in the wavenumber domain are located in a sub-space representing the LoS components. 
Thus, it is reasonable to describe the channel using that low-dimensional sub-space in the wavenumber domain while omitting the NLoS counterpart.
Specifically, based on the full-size WTMs, i.e., $\mathbf{\Phi }_{\mathrm{U}}$ and $\mathbf{\Phi }_{\mathrm{B}}$, two truncated WTMs with lower dimensions need to be obtained, via which the LoS sub-space of ${\tilde{\mathbf{H}}}_{\rm a}$ can be extracted.  
In \cite{tang2023line}, the boundaries of the LoS sub-space are derived based on the array geometry, meaning that the precise locations of transceivers are assumed to be known.
However, when the location information is missing at transceivers, the method cannot be used.
Therefore, we propose a sensing method to estimate the boundaries of the subspace.

\subsubsection{Downlink Sensing}
In the DL sensing phase, the transmit beamformer $\mathbf{p}_t$ at the BS is designed as
\begin{align}
	\mathbf{p}_t=\frac{1}{\sqrt{N}}\mathbf{\Phi }_{\mathrm{B}}^{}\mathbf{c}_{t}^{\rm DL}\oslash \left( \mathbf{\Phi }_{\mathrm{B}}^{}\mathbf{c}_{t}^{\rm DL} \right)^{|\cdot|},
\end{align}
where $\mathbf{c}_{t}^{\rm DL} \in \mathbb{R}^{N \times 1}$ is a constant vector.
When ${\mathbf{y}}_t^{\mathrm{DL}}$ is received at the UE via DL, the received gain vector in the wavenumber domain can be obtained by $\mathbf{w}_t^{\rm DL}= \mathbf{\Phi }_{\mathrm{U}}^{H}\mathbf{y}_{t}^{\mathrm{DL}}$.
To cancel out the interference caused by noise, we average $\mathbf{w}_t^{\rm DL}$ over rounds.
Specifically, let $K=T_{\rm s}$ be the number of pilots sent by the BS via DL, the averaged gain vector at the UE can be expressed as $\hat{\mathbf{w}}^{\mathrm{DL}}=\frac{1}{K}( \sum\nolimits_{t=0}^{K-1}{\mathbf{w}_{t}^{\mathrm{DL}}} ) ^{|\cdot |}
$.
With $\hat{\mathbf{w}}^{\mathrm{DL}}$, the boundaries of the LoS sub-space on the UE side are given by
\begin{align}
	i_{\max}^{\left( \mathrm{e} \right)}&=\mathrm{arg}\max _{i\in \mathcal{G} _{\boldsymbol{\kappa }}^{}}\left( \left[ \hat{\mathbf{w}}^{\mathrm{DL}} \right] _i>\Gamma_{{\bf{w}, \rm DL}} \right), \label{eq:i_max} \\
	i_{\min}^{\left( \mathrm{e} \right)}&=\mathrm{arg}\min _{i\in \mathcal{G} _{\boldsymbol{\kappa }}^{}}\left( \left[ \hat{\mathbf{w}}^{\mathrm{DL}} \right] _i>\Gamma_{{\bf{w}, \rm DL}} \right), \label{eq:i_min}
\end{align}
where $\Gamma_{{\bf{w}, \rm DL}}$ denotes the predefined threshold for the DL sensing.
Based on the above boundaries, a sub-set of $\mathcal{G} _{\boldsymbol{\kappa }}$, representing the LoS links, can be expressed as $\mathcal{G} _{\boldsymbol{\kappa }}^{\left( \mathrm{e} \right)} =\{i\in \mathbb{Z} :i_{\min}^{\left( \mathrm{e} \right)}\leqslant  i \leqslant i_{\max}^{\left( \mathrm{e} \right)} \}$.
Correspondingly, by extracting the columns of $\mathbf{\Phi }_{\mathrm{U}}$, that are indexed by $\mathcal{G} _{\boldsymbol{\kappa }}^{\left( \mathrm{e} \right)}$, the truncated WTM on the UE side can be expressed as
\begin{align}
	\mathbf{\Phi }_{\mathrm{U}}^{\left( \mathrm{e} \right)}=\left[ \boldsymbol{\phi }_{\mathrm{U},i_{\min}^{\left( \mathrm{e} \right)}}^{},...,\boldsymbol{\phi }_{\mathrm{U},i_{\max}^{\left( \mathrm{e} \right)}}^{} \right] \in \mathbb{C} ^{M\times \left| \mathcal{G} _{\boldsymbol{\kappa }}^{\left( \mathrm{e} \right)} \right|}. \label{eq:trans_U_eq}
\end{align} 
With the above, we can obtain a direct mapping from the space domain to the truncated wavenumber domain on the UE side.

\subsubsection{Uplink Sensing}
In the UL sensing phase, the receive beamformer $\mathbf{s}_t$ at the UE can be expressed as
\begin{align}
	\mathbf{s}_t=\frac{1}{\sqrt{M}}\mathbf{\Phi }_{\mathrm{U}}^{}\mathbf{c}_{t}^{\rm UL}\oslash ( \mathbf{\Phi }_{\mathrm{U}}^{}\mathbf{c}_{t}^{\rm UL} )^{|\cdot|},
\end{align}	
where $\mathbf{c}_{t}^{\rm UL} \in \mathbb{R}^{N \times 1}$ is a constant vector.
Once ${\mathbf{y}}_t^{\mathrm{UL}}$ is received at the BS, the received gain vector in the wavenumber domain can be obtained by $\mathbf{w}_t^{\rm UL}=\mathbf{\Phi }_{\mathrm{B}}^{T}\mathbf{y}_{t}^{\mathrm{UL}}$.
Then, given that the number of pilots sent by the UE is $K=T_{\rm s}$, the averaged received gain vector at the BS is given by $\hat{\mathbf{w}}^{\mathrm{UL}}=\frac{1}{K}( \sum\nolimits_{t=0}^{K-1}{\mathbf{w}_{t}^{\mathrm{UL}}} ) ^{|\cdot |}$. 
With $\hat{\mathbf{w}}^{\mathrm{UL}}$, the boundaries of the LoS sub-space on the BS side can be defined by
\begin{align}
	j_{\max}^{\left( \mathrm{e} \right)}&=\mathrm{arg}\max _{j\in \mathcal{G} _{\boldsymbol{k }}^{}}\left( \left[ \hat{\mathbf{w}}^{\mathrm{UL}} \right] _j>\Gamma_{{\bf{w}, \rm UL}} \right), \label{eq:j_max}\\
	j_{\min}^{\left( \mathrm{e} \right)}&=\mathrm{arg}\min _{j\in \mathcal{G} _{\boldsymbol{k }}^{}}\left( \left[ \hat{\mathbf{w}}^{\mathrm{UL}} \right] _j>\Gamma_{{\bf{w}, \rm UL}} \right), \label{eq:j_min}
\end{align}
where $\Gamma_{{\bf{w}, \rm UL}}$ is the pre-defined threshold for estimation in UL.
Similarly, by extracting the columns of $\mathbf{\Phi }_{\mathrm{B}}$ indexed by $\mathcal{G} _{\boldsymbol{k}}^{\left( \mathrm{e} \right)} =\{j\in \mathbb{Z} :j_{\min}^{\left( \mathrm{e} \right)}\leqslant  j \leqslant j_{\max}^{\left( \mathrm{e} \right)} \}$, the truncated WTM on the BS side can be expressed as
\begin{align}
	 \mathbf{\Phi }_{\mathrm{B}}^{\left( \mathrm{e} \right)}=\left[ \boldsymbol{\phi }_{\mathrm{B},j_{\min}^{\left( \mathrm{e} \right)}}^{},...,\boldsymbol{\phi }_{\mathrm{B},j_{\max}^{\left( \mathrm{e} \right)}}^{} \right] \in \mathbb{C} ^{N\times \left| \mathcal{G} _{\mathbf{k}}^{\left( \mathrm{e} \right)} \right|}.
	  \label{eq:trans_B_eq}
\end{align}

It is noted that compared to the full-size WTMs, i.e., $\mathbf{\Phi }_{\mathrm{U}}$ and $\mathbf{\Phi }_{\mathrm{B}}$, the truncated ones have much lower dimensions, i.e., $|\mathcal{G} _{\mathbf{k}}^{\left( \mathrm{e} \right)} |\ll |\mathcal{G} _{\mathbf{k}} |$ and  $|\mathcal{G} _{\boldsymbol{\kappa }}^{\left( \mathrm{e} \right)} | \ll |\mathcal{G} _{\boldsymbol{\kappa }}|$. 
Using $\mathbf{\Phi }_{\mathrm{U}}^{\left( \mathrm{e} \right)}$ and $\mathbf{\Phi }_{\mathrm{B}}^{\left( \mathrm{e} \right)}$, the channel can be expressed by
\begin{align}
	 \mathbf{H} \approx \sqrt{MN}\mathbf{\Phi }_{\mathrm{U}}^{\left( \mathrm{e} \right)}\tilde{\mathbf{H}}_{\mathrm{e}}\left( \mathbf{\Phi }_{\mathrm{B}}^{\left( \mathrm{e} \right)} \right) ^H. \label{eq:H_e}
 \end{align}
Similarly, we have $\tilde{\mathbf{H}}_{\mathrm{e}} \approx \frac{1}{\sqrt{MN}}( \mathbf{\Phi }_{\mathrm{U}}^{\left( \mathrm{e} \right)} ) ^H\mathbf{H\Phi }_{\mathrm{B}}^{\left( \mathrm{e} \right)}$. 
Finally, the procedures are summarized in {\textbf{Algorithm 1}}. \footnote{It is noted that our method can be generalized to the non-parallel case since the boundaries are obtained by sensing. }
The approximation error of \eqref{eq:H_e} is attributed to two primary factors: 1)~the substitution of the integration in \eqref{eq:ft_cont} with a summation of finite terms in \eqref{eq:H_discrete}, and 2)~the choice of different sensing thresholds, denoted by $\Gamma_{\bf{w}}$.
Given that the ELAAs are electromagnetically large, the approximation error incurred by the former factor is negligible, as highlighted by \cite{pizzo2020holographic} and \cite{pizzo2021fourier}. 
Concerning the latter factor, the approximation error tends to increase with a larger $\Gamma_{\bf{w}}$.
Specifically, setting a larger $\Gamma_{\bf{w}}$ will result in a smaller truncated wavenumber-domain subspace while leading to a reduction in channel power.

\begin{algorithm}[t!] 
	\caption{Sensing Phase of Proposed STT Scheme}
	\begin{algorithmic}[1]  \small
		\STATE{\textbf{Initialization}: 
			Initialize $\mathbf{\Phi }_{\mathrm{U}}^{}$ and $\mathbf{\Phi }_{\mathrm{B}}^{}$, the maximum training round in the sensing phase $T_{\rm s}$, $K={T_{\mathrm{s}}} $, and $\Gamma_{{\bf{w}, \rm DL}}$ and $\Gamma_{{\bf{w}, \rm UL}}$;
			initialize ${\mathbf{c}_{0}^{\rm UL}}$ and ${\mathbf{c}_{0}^{\rm DL}}$ as constant vectors.
			Let $t=0$ to start the sensing phase.}
		\FOR{$t = 0, 1, ..., T_{\rm s}$}
		\STATE{\textbf{BS}: Transmit $\mathbf{p}_t=\frac{1}{\sqrt{N}}\mathbf{\Phi }_{\mathrm{B}}^{}\mathbf{c}_{t}^{\rm DL}\oslash \left( \mathbf{\Phi }_{\mathrm{B}}^{}\mathbf{c}_{t}^{\rm DL} \right)^{|\cdot|}$.}
		\STATE{\textbf{UE}: Receive $\mathbf{y}_{t}^{\mathrm{DL}}$ and obtain $\mathbf{w}_t^{\rm DL}= \mathbf{\Phi }_{\mathrm{U}}^{H}\mathbf{y}_{t}^{\mathrm{DL}}$. }
		\STATE{\textbf{UE}: Transmit $\mathbf{s}_t=\frac{1}{\sqrt{M}}\mathbf{\Phi }_{\mathrm{U}}^{}\mathbf{c}_{t}^{\rm UL}\oslash ( \mathbf{\Phi }_{\mathrm{U}}^{}\mathbf{c}_{t}^{\rm UL} )^{|\cdot|}$.}
		\STATE{\textbf{BS}: Receive $\mathbf{y}_{t}^{\mathrm{UL}}$ and obtain $\mathbf{w}_t^{\rm UL}= \mathbf{\Phi }_{\mathrm{B}}^{T}\mathbf{y}_{t}^{\mathrm{UL}}$. }
		\ENDFOR
		\STATE \textbf{UE}: obtain $\hat{\mathbf{w}}^{\mathrm{DL}}=\frac{1}{K}( \sum\nolimits_{t=0}^{K-1}{\mathbf{w}_{t}^{\mathrm{DL}}} ) ^{|\cdot |}$ and  $\mathbf{\Phi }_{\mathrm{U}}^{\left( \mathrm{e} \right)}$ by \eqref{eq:i_max} and \eqref{eq:i_min}.
		\STATE \textbf{BS}: obtain $\hat{\mathbf{w}}^{\mathrm{UL}}=\frac{1}{K}( \sum\nolimits_{t=0}^{K-1}{\mathbf{w}_{t}^{\mathrm{UL}}} ) ^{|\cdot |}$ and  $\mathbf{\Phi }_{\mathrm{B}}^{\left( \mathrm{e} \right)}$ by \eqref{eq:j_max} and \eqref{eq:j_min}.
	\end{algorithmic}
\end{algorithm}

\subsection{Training Phase}
\begin{figure*}[t!]
	\centering
	\includegraphics[width=1\linewidth]{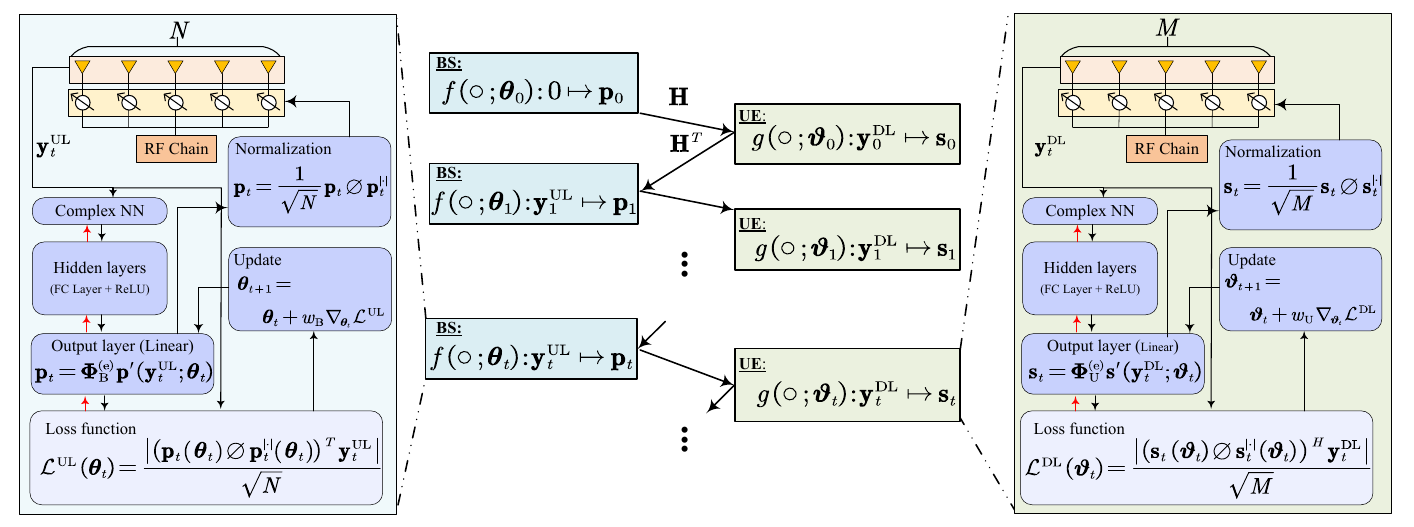}
	\caption{An overview the of proposed STT method for the single-beam case. Data/gradient flows are denoted by the black/red line.}
	\label{fig:SD_NN_struct}
\end{figure*}
With the sensing results, i.e., $\mathbf{\Phi }_{\mathrm{U}}^{\left( \mathrm{e} \right)}$ and $\mathbf{\Phi }_{\mathrm{B}}^{\left( \mathrm{e} \right)}$, and the approximation in \eqref{eq:H_e}, the objective function of problem \eqref{eq:SD-Obj} can be rewritten as follows:
\begin{align}
	| \mathbf{s}^H\mathbf{Hp} | ^2 \approx | \left( \mathbf{s}^{\prime} \right) ^H\tilde{\mathbf{H}}_{\mathrm{e}}\mathbf{p}^{\prime} | ^2, \label{eq:obj-SD}
\end{align}
where $\mathbf{p}=\mathbf{\Phi }_{\mathrm{B}}^{\left( \mathrm{e} \right)}\mathbf{p}^{\prime}
$ and $\mathbf{s}= \mathbf{\Phi }_{\mathrm{U}}^{\left( \mathrm{e} \right)}\mathbf{s}^{\prime}$.
The approximation error of \eqref{eq:obj-SD} is incurred by mapping the space-domain channel into the truncated wavenumber domain as described by \eqref{eq:H_e}.
In this case, low-dimensional ${\bf p}^\prime$ and ${\bf s}^\prime$ can be optimized according to $\tilde{\mathbf{H}}_{\mathrm{e}}$, which can simplify the beam training problem.  
In the sequel, the active-sensing-based training algorithms in the DL and UL are elaborated. 

\subsubsection{Downlink Training}
In the DL training, the objective for the UE is to find a receive beamformer $\mathbf{s}_t \in \mathbb{C}^{M \times 1}$ to produce the highest beam gain, which is defined as $U_{\rm S}^{\rm DL}\left( \mathbf{s}_t\right) =| \mathbf{s}_t^H\mathbf{Hp}_t | ^2 \approx | \mathbf{s}_t^H{\mathbf{y}}_t^{\mathrm{DL}} | ^2$, where ${\mathbf{y}}_t^{\mathrm{DL}}$ can be seen as a noisy observation of $\mathbf{Hp}_t$.
The resulting optimization problem for DL training is given by
\begin{subequations}
		\vspace{-1mm}
	\begin{align} 
		&\max_{\mathbf{s}_t}~U_{\rm S}^{\rm DL}\left( \mathbf{s}_t\right)  \label{eq:SD-Obj-DL}\\ 
		&~{\rm{s.t.}}~  \eqref{eq:P-c2}. \notag
	\end{align}
\end{subequations}
The above problem is generally challenging to solve due to the non-convexity caused by \eqref{eq:P-c2}, and the high dimensions of ${\bf s}_t$ caused by the density of antennas.
To address these challenges, we first find a low-dimensional $\mathbf{s}^\prime$ in the wavenumber domain based on the received signal.
Then, we employ $\mathbf{\Phi }_{\mathrm{U}}^{\left( \mathrm{e} \right)}$ to map $\mathbf{s}^\prime$ back to the space domain.
To this end, a neural network (NN) is exploited, named as the UE-NN module.
Specifically, the UE-NN module can be seen as a mapping function, described by $g\left( \circ ;\boldsymbol{\vartheta }_t \right) :\mathbf{y}_t^{\text{DL}} \mapsto \mathbf{s}_{t}$, where $\boldsymbol{\vartheta}$ is the vector composed by all the trainable parameters. 
The content of this module is a matrix production between two sub-modules, i.e., a complex NN mapping function and a WTM, described by
\begin{align}
	\mathbf{s}_{t+1}\left( \boldsymbol{\vartheta }_t \right) =g\left( \mathbf{y}_t^{\text{DL}} ;\boldsymbol{\vartheta }_t \right) = \mathbf{\Phi }_{\mathrm{U}}^{\left( \mathrm{e} \right)}{\mathbf{s}}^{\prime}\left( \mathbf{y}_t^{\text{DL}};\boldsymbol{\vartheta }_t \right), \label{eq:ue_mapping}
\end{align}
where ${\mathbf{s}}^{\prime}\left( \mathbf{y}_t^{\text{DL}}; \boldsymbol{\vartheta }_t \right): \mathbb{C}^{M\times1} \mapsto \mathbb{C}^{| \mathcal{G} _{\boldsymbol{\kappa }}^{\left( \mathrm{e} \right)} | \times 1}$ denotes the complex NN mapping function and $\mathbf{\Phi }_{\mathrm{U}}^{\left( \mathrm{e} \right)}: \mathbb{C}^{| \mathcal{G} _{\boldsymbol{\kappa }}^{\left( \mathrm{e} \right)} |\times 1}\mapsto \mathbb{C}^{M \times 1}$ denotes the truncated WTM on the UE side. 
It is noted that the computation complexity is most attributed to the training of ${\mathbf{s}}^{\prime}\left( \mathbf{s}_t; \boldsymbol{\vartheta }_t \right)$, since $\mathbf{\Phi }_{\mathrm{U}}^{\left( \mathrm{e} \right)}$ can be obtained in $\mathcal{O}(1)$ complexity by sensing.
To satisfy \eqref{eq:P-c2}, vector $\mathbf{s}_{t}$ is normalized according to $\mathbf{s}_{t}=\frac{1}{\sqrt{M}}\mathbf{s}_{t}\oslash \mathbf{s}_{t}^{\left| \cdot \right|}$.
According to \eqref{eq:SD-Obj-DL}, the loss function and the update rule of the UE-NN module are given by
\begin{align}
	\mathcal{L}^{\mathrm{DL}} \left( \boldsymbol{\vartheta }_t \right) &=\frac{1}{\sqrt{M}}\left| \left( \mathbf{s}_{t}^{}\left( \boldsymbol{\vartheta }_t \right) \oslash \mathbf{s}_{t}^{\left| \cdot \right|}\left( \boldsymbol{\vartheta }_t \right) \right) ^H\mathbf{y}_{t}^{\mathrm{DL}} \right|,\\ \label{loss-ue} 
	\boldsymbol{\vartheta }_{t+1}&=\boldsymbol{\vartheta }_t+w_{\mathrm{U}}\nabla _{\boldsymbol{\vartheta }_t} \mathcal{L}^{\mathrm{DL}},
\end{align}
where $w_{\mathrm{U}}$ is the learning rate on the UE side.
It is noted that since we are maximizing the loss function, gradient ascent is utilized to maximize the objective.

\subsubsection{Uplink Training}
In the UL training, the objective for the BS is to find a transmit beamformer $\mathbf{p}_t \in \mathbb{C}^{N \times 1}$ to produce the highest beam gain, which is defined as $U_{\rm S}^{\rm UL}\left( \mathbf{p}_t\right) =|\mathbf{p}_{t}^{T}\mathbf{H}^T\mathbf{s}_{t}^{*}|^2\approx |\mathbf{p}_{t}^{T}\mathbf{y}_{t}^{\mathrm{UL}}|^2$, where ${\mathbf{y}}_t^{\mathrm{UL}}$ can be seen as a noisy observation of $\mathbf{H}^T \mathbf{s}_{t}^{*}$.
In contrast to the DL case, the UE transmits pilots using a conjugate beamformer, i.e., $\mathbf{s}_{t}^{*}$, in the UL.
Therefore, the resulting optimization problem for the UL is given by
\begin{subequations}
	\begin{align} 
		&\max_{\mathbf{p}_t}~U_{\rm S}^{\rm UL}\left( \mathbf{p}_t\right)  \label{eq:SD-Obj-UL}\\ 
		&~{\rm{s.t.}}~ \eqref{eq:P-c1}. \notag
	\end{align}
\end{subequations}
Similar to the DL case, an NN is used to find $\mathbf{p}^\prime$ in the wavenumber domain, which is then converted to the space domain via $\mathbf{\Phi }_{\mathrm{B}}^{\left( \mathrm{e} \right)}$. 
Specifically, the BS-NN module can be expressed as $f\left( \circ ;\boldsymbol{\theta }_t \right) :\mathbf{y}_t^{\text{UL}} \mapsto \mathbf{p}_{t}$, with $\boldsymbol{\theta}$ is the trainable parameter vector.
The whole module can be seen as a matrix production of two sub-modules, which can be described by
\begin{align}
	\mathbf{p}_{t}(\boldsymbol{\theta }_t)=f\left( \mathbf{y}_t^{\text{UL}} ;\boldsymbol{\theta }_t \right)  = \mathbf{\Phi }_{\mathrm{B}}^{\left( \mathrm{e} \right)}{\mathbf{p}}^{\prime}\left( \mathbf{y}_t^{\text{UL}};\boldsymbol{\theta }_t \right), \label{eq:bs_mapping}
\end{align}
where ${\mathbf{p}}^{\prime}\left( \circ;\boldsymbol{\theta }_t \right): \mathbb{C}^{N\times1} \mapsto \mathbb{C}^{| \mathcal{G} _{\boldsymbol{k }}^{\left( \mathrm{e} \right)} | \times 1}$ denotes the complex NN mapping function and $
\mathbf{\Phi }_{\mathrm{B}}^{\left( \mathrm{e} \right)}: {\mathbb{C}}^{|{\mathcal{G} _{k}^{\left( \mathrm{e} \right)}} | \times 1} \mapsto {\mathbb{C}}^{N\times 1}$ denotes the truncated WTM at the BS. 
Furthermore, to satisfy \eqref{eq:P-c1}, the transmit beamformer $\mathbf{p}_{t}(\boldsymbol{\theta }_t)$ is normalized according to $\mathbf{p}_{t}=\frac{1}{\sqrt{N}}\mathbf{p}_{t}\oslash \mathbf{p}_{t}^{\left| \cdot \right|}$.
Based on \eqref{eq:SD-Obj-UL}, the loss function and the update rule of the BS-NN module are given by
\begin{align}
	\mathcal{L}^{\mathrm{UL}} \left( \boldsymbol{\theta }_t \right) &=\frac{1}{\sqrt{N}}\left| \left( \mathbf{p}_{t}^{}(\boldsymbol{\theta }_t)\oslash \mathbf{p}_{t}^{\left| \cdot \right|}(\boldsymbol{\theta }_t) \right) ^T\mathbf{y}_{t}^{\mathrm{UL}} \right|,
	\label{loss-bs} \\
	\boldsymbol{\theta }_{t+1}&=\boldsymbol{\theta }_t+w_{\mathrm{B}}\nabla _{\boldsymbol{\theta }_t}\mathcal{L}^{\mathrm{UL}}, \label{eq:wei_up_UL}
\end{align}
where $w_{\mathrm{B}}$ is the learning rate on the BS side.

The active-sensing-based method is illustrated by Fig. \ref{fig:SD_NN_struct}, and then summarized in \textbf{Algorithm 2}.
To initialize the training process, at the initial round, i.e., $t=0$, the BS NN module can be fed with any suitable vectors to continue the UL transmission.
In our case, an all-zero vector is fed as the initial input.

\textbf{Remark 1:} It can be seen from \textbf{Algorithm 2} that the training of NNs is conducted online.
	Different from conventional batch-based offline learning, the NN models are trained incrementally as each new data point arrives, i.e., online learning, which is also called online machine learning.
	Three major factors drive the usage of this method: 1)~the dimensions of truncated WTMs at transceivers may vary according to the sensing results. This variability prevents us to determine the output dimensions in advance during the offline training stage, thus necessitating an adaptive approach; 2)~the NN model is updated continuously, which enables the transceivers to adapt to new patterns in the received signals as the ping-pong process goes on; and 3)~due to the rank-deficient structures of near-field channels, the truncated wavenumber-domain channel representations are of low dimensions, which makes the online learning feasible and practical in terms of computational complexity.

\begin{algorithm}[t!] 
	\caption{Training Phase of Single-beam STT Scheme}
	\label{alg:narrow_active_sensing}
	\begin{algorithmic}[1] \small
		\STATE{\textbf{Initialization}: obtain $\mathbf{\Phi }_{\mathrm{B}}^{\left( \mathrm{e} \right)}$ and $\mathbf{\Phi }_{\mathrm{U}}^{\left( \mathrm{e} \right)}$ via {\textbf{Algorithm 1}};
			initialize $\boldsymbol{\vartheta }_0$ and $\boldsymbol{\theta }_0$ and the maximum training round $T_{\rm a}$;
			obtain $\mathbf{p}_0$ by feeding the BS with a zero vector; set learning rates $w_{\rm B}$ and $w_{\rm U}$.}
		\FOR{$t = 0, 1, 2, ..., T_{\rm a}$}
		\STATE{\textbf{UE}: {\makecell[l]{a)~obtain the receive beamformer by $\mathbf{s}_{t} = g\left( \mathbf{y}_t^{\text{DL}} ;\boldsymbol{\vartheta }_t \right)$; \\b)~obtain $\boldsymbol{\vartheta }_{t+1}$ by updating $\boldsymbol{\vartheta }_{t}$ using \eqref{loss-ue}; \\c)~transmit $c_{\text{U}}$ using $\mathbf{s}_{t}^*$.}}}
		\STATE{\textbf{BS}: {\makecell[l]{a)~obtain the transmit beamformer by $\mathbf{p}_{t}=f\left( \mathbf{y}_{t}^{\mathrm{UL}} ;\boldsymbol{\theta }_t \right)$; \\ b)~obtain $\boldsymbol{\theta }_{t+1}$ by updating  $\boldsymbol{\theta }_{t}$ using \eqref{eq:wei_up_UL}; \\ c)~transmit $c_{\text{B}}$ using $\mathbf{p}_{t}$.}}}
		\ENDFOR
	\end{algorithmic}
\end{algorithm}

\subsection{Stability, Required Information, and Cost Analysis} 
\subsubsection{Stability} The stability of the proposed STT scheme relies on the channel condition. 
Intuitively, when the channel condition is poor, the proposed STT scheme would have difficulty facilitating beam training based on the noisy observations of pilots. 
Nevertheless, in practice, the high-frequency channel is dominated by the LoS component, which can ensure a high signal-to-noise ratio (SNR) at the receiver.

\subsubsection{Required Information} 
For the single-beam case, the truncated WTMs at the transceivers, i.e., $\mathbf{\Phi }_{\mathrm{B}}^{\left( \mathrm{e} \right)}$ and $\mathbf{\Phi }_{\mathrm{U}}^{\left( \mathrm{e} \right)}$, are needed.
Referring to \eqref{eq:H_discrete}, \eqref{eq:H_discrete-1}, \eqref{eq:H_discrete-2}, and  \eqref{eq:H_discrete-3}, the WTMs can be constructed locally at transceivers by sampling the wavenumber domain.
Then, the truncated WTMs can be obtained via the sensing phase. 
Thanks to the sensing phase, this information can be obtained without explicit information exchange between transceivers.

\subsubsection{Cost} \label{sect:computaional_complexity_single_beam}
 By adopting the STT scheme, beam training can be carried out in the truncated low-dimensional wavenumber domain.
	The primary cost of the single-beam STT arises from its computational complexity.
	Specifically, for the sensing phase, the computation complexity of obtaining original WTMs, i.e., $\mathbf{\Phi }_{\mathrm{U}}$ and $\mathbf{\Phi }_{\mathrm{B}}$, is $\mathcal{O}(1)$.
	Then, the computational complexity to obtain the truncated WTMs, i.e., $\mathbf{\Phi }_{\mathrm{U}}^{\left( \mathrm{e} \right)}$ and $\mathbf{\Phi }_{\mathrm{B}}^{\left( \mathrm{e} \right)}$, is also $\mathcal{O}(1)$, as it mainly involves averaging the received pilots.
	For the training phase, let $L_{\rm B}$ and $o_{\rm B}^{l}$ be the number of hidden layers of the BS-NN module and the number of neurons in $l$-th layer, respectively.
	The input layer has a dimension of $N$ and the output layer has a dimension of $|\mathcal{G} _{\mathbf{k}}^{\left( \mathrm{e} \right)}|$.
	Then, the number of weights at the input and output layers can be respectively expressed as $No_{\rm B}^{1}$ and $|\mathcal{G} _{\mathbf{k}}^{\left( \mathrm{e} \right)}|o_{\rm B}^{L_{\rm B}}$.
	Hence, the total number of weights that necessitate updating is given by $No_{\rm B}^{1} + |\mathcal{G} _{\mathbf{k}}^{\left( \mathrm{e} \right)}|o_{\rm B}^{L_{\rm B}} + \sum\nolimits_{l=2}^{L_{\mathrm{B}}}{o_{\mathrm{B}}^{l-1}}o_{\mathrm{B}}^{l}$.
	Letting $J$ be the computational complexity of training a weight, the total computational complexity to train the NN at the BS can be expressed as $\mathcal{O}(JT_{\rm a}(No_{\rm B}^{1} + |\mathcal{G} _{\mathbf{k}}^{\left( \mathrm{e} \right)}|o_{\rm B}^{L_{\rm B}} + \sum\nolimits_{l=2}^{L_{\mathrm{B}}}{o_{\mathrm{B}}^{l-1}}o_{\mathrm{B}}^{l}))$. 
	For the UE, the computational complexity can be obtained in a similar way, which can be expressed as $\mathcal{O}(JT_{\rm a}(Mo_{\rm U}^{1} + |\mathcal{G} _{\boldsymbol{\kappa}}^{\left( \mathrm{e} \right)}|o_{\rm U}^{L_{\rm U}} + \sum\nolimits_{l=2}^{L_{\mathrm{U}}}{o_{\mathrm{U}}^{l-1}}o_{\mathrm{U}}^{l}))$, where $L_{\rm U}$ and $o_{\rm U}^{l}$ represent the number of layers and the number of neurals of the $l$-th layer of the UE NN module, respectively.
	It is important to note that by leveraging the dominance of LoS channel, we have $|\mathcal{G} _{\mathbf{k}}^{\left( \mathrm{e} \right)}| \ll N$ and $|\mathcal{G} _{\boldsymbol{\kappa}}^{\left( \mathrm{e} \right)}|\ll M$, which reduce the number of trainable parameters.
	Besides, with the parallel computation on graphics processing units (GPUs), these parameters can be trained quickly and efficiently.

\section{STT for Multi-Beam Cases}\label{sect:MDS}
For the multi-beam case, the SE is specified by \eqref{eq:se}, which can be maximized by choosing the singular values of $\mathbf{H}$.
Notably, in the beam training phase, our objective is to find the optimal beamformers $\mathbf{S}$ and $\mathbf{P}$, while the optimal $\mathbf{\Lambda }$ can be found using water-filling during data transmission. 
Moreover, according to \cite{sohrabi2022active} and \cite{yu2016alternating}, we have $\mathbf{C}\approx \mathbf{I}_{N_s}$ for the optimal beamformers, meaning the columns of beamformers are orthogonal to each other.
Therefore, problem \eqref{eq:P} can be reformulated as:
\begin{subequations}
	\vspace{-0.1cm}
	\begin{align} 
		&\max_{\mathbf{S},\mathbf{P}} ~\left| \mathbf{S}_{}^{H}\mathbf{HP\Lambda \Lambda }^H\mathbf{P}_{}^{H}\mathbf{H}^H\mathbf{S}_{} \right| 
		\label{eq:P-MD}\\ 
		&~{\rm{s.t.}}~ \eqref{eq:P-c1}~{\rm and }~ \eqref{eq:P-c2} . \notag
	\end{align}
\end{subequations}
It can be seen that when $\mathbf{S}$ and $\mathbf{P}$ equal the left and the right singular vectors of $\mathbf{H}$, the objective of problem \eqref{eq:P-MD} is maximized, while the SE is maximized simultaneously.
To solve the above problem, a multi-beam STT is proposed, in which the beams in $\mathbf{S}$ or $\mathbf{P}$ are trained successively, while a Gram-Schmidt method is utilized to guarantee the orthogonality among beams. 

\subsection{Signal Model and Transmission Protocol}
Similar to the single-beam case, ping-pong pilots are utilized.
The received signal at the UE via DL and that at the BS via UL are given by
\begin{align}
	\mathbf{y}_{t}^{\mathrm{DL}}&=\mathbf{HP}_t\mathbf{\Lambda }_{\mathrm{B}, t}\mathbf{c}_{\mathrm{B},t}+\mathbf{n}_{\mathrm{U}}, \\
	\mathbf{y}_{t}^{\mathrm{UL}}&=\mathbf{H}^T\mathbf{S}_{t}^{*}\mathbf{\Lambda }_{\mathrm{U}, t}\mathbf{c}_{\mathrm{U},t}+\mathbf{n}_{\mathrm{B}},
\end{align}
where $\mathbf{c}_{\mathrm{B},t} \in \mathbb{C}^{N_{\rm s} \times 1}$ and $\mathbf{c}_{\mathrm{U},t}\in \mathbb{C}^{N_{\rm s} \times 1}$ denote the transmitted pilot signals at the BS and the UE, respectively, $\mathbf{\Lambda }_{\mathrm{B}, t} \in \mathbb{R}^{N_{\rm s} \times N_{\rm s}}$ and $\mathbf{\Lambda }_{\mathrm{U}, t} \in \mathbb{R}^{N_{\rm s} \times N_{\rm s}}$ are diagonal power allocation matrices during beam training, whose entries satisfy $\mathrm{Tr}\left\{ \mathbf{\Lambda }_{\mathrm{B}, t}^{T}\mathbf{\Lambda }_{\mathrm{B}, t} \right\} =P_{\mathrm{B}}
$ and $\mathrm{Tr}\left\{ \mathbf{\Lambda }_{\mathrm{U}, t}^{T}\mathbf{\Lambda }_{\mathrm{U}, t} \right\} =P_{\mathrm{U}}$, respectively, and $\mathbf{n}_{\mathrm{U}} \sim \mathcal{C} \mathcal{N} (\boldsymbol{0},\sigma^{2}\mathbf{I}_M)
$ and $ {\mathbf{n}_{\mathrm{B}}}\sim \mathcal{C} \mathcal{N} (\boldsymbol{0},\sigma^{2}\mathbf{I}_N)$ are the complex Gaussian noise at the UE and the BS, respectively.

\begin{figure}
	\centering
	\includegraphics[width=1\linewidth]{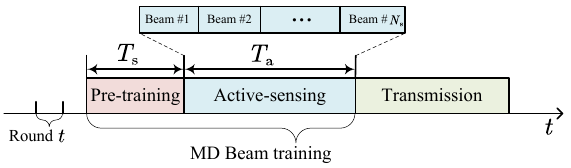}
	\caption{An illustration of the multi-beam STT scheme.} \label{fig:STT_multi-beam}
\end{figure}

Like the single-beam case, the beam training protocol is divided into two phases, i.e., a sensing phase lasting for $T_{\rm s}$ and a training phase lasting for $T_{\rm a}$. 
However, different from the single-beam STT, the multi-beam STT trains the beams successively during the training phase.
For instance, when the $i$-th beam is trained sufficiently, the successive beam indexed by $i+1$ will be trained in the space that is orthogonal to all the predecessor beams.
It is noted that the duration for training each beam is not necessarily identical and is controlled by a threshold.
Similar to the single-beam case, the overall procedure of the multi-beam STT scheme is illustrated by Fig. \ref{fig:STT_multi-beam}, where ``MD" refers to ``multiple data streams".
It is important to highlight that the sensing phase is carried out only once, since all beams are sharing one common wavenumber domain. 
\subsection{Sensing Phase}
Regardless of the single-beam or the multi-beam cases, the function of sensing is to obtain the truncated WTMs.
Hence, in the multi-beam case, the sensing phase is similar to that of the single-beam case and can be realized via \textbf{Algorithm 1}.
The only difference lies in that power is allocated uniformly, i.e., $\mathbf{\Lambda }_{\mathrm{B},t}=\sqrt{\frac{P_{\mathrm{B}}}{N_{\mathrm{s}}}}\mathbf{I}_{N_{\mathrm{s}}}$ and $\mathbf{\Lambda }_{\mathrm{U},t}=\sqrt{\frac{P_{\mathrm{U}}}{N_{\mathrm{s}}}}\mathbf{I}_{N_{\mathrm{s}}}$. 
It is important to point out that although the beams are trained in a one-by-one fashion, only one sensing phase is needed.

\subsection{Training Phase} \label{Sect:training_MD}
With the sensing results, i.e., $\mathbf{\Phi }_{\mathrm{U}}^{\left( \mathrm{e} \right)}$ and $\mathbf{\Phi }_{\mathrm{B}}^{\left( \mathrm{e} \right)}$, the objective \eqref{eq:P-MD} can be converted to the low-dimensional wavenumber domain via
\begin{align}
	\eqref{eq:P-MD} \approx \left| \left( \mathbf{S}^{\prime} \right) ^H\tilde{\mathbf{H}}_{\mathrm{e}}\mathbf{P}^{\prime}\mathbf{\Lambda \Lambda }^H\left( \mathbf{P}^{\prime} \right) ^H\tilde{\mathbf{H}}_{\mathrm{e}}^{H}\mathbf{S}^{\prime} \right|,
\end{align}
where $\mathbf{S}=\mathbf{\Phi }_{\mathrm{U}}^{\left( \mathrm{e} \right)}\mathbf{S}^{\prime}$ and $\mathbf{P}=\mathbf{\Phi }_{\mathrm{B}}^{\left( \mathrm{e} \right)}\mathbf{P}^{\prime}$.
In the sequel, the training methods in the DL and UL phases are elaborated. 

\begin{figure*}[!ht]
	\centering
	\includegraphics[width=0.85\linewidth]{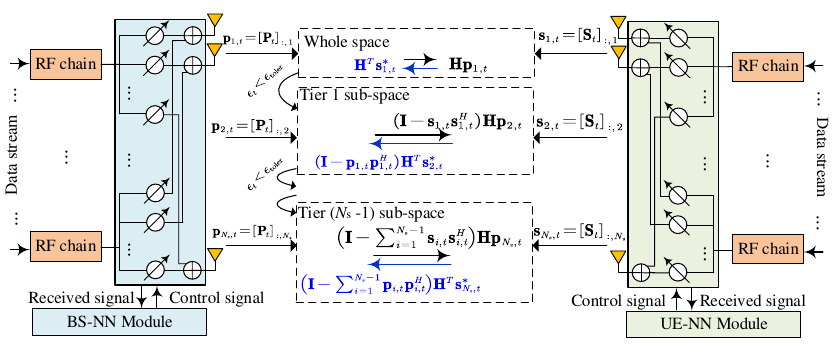}
	\caption{An overview of the proposed STT method for the multi-beam case.}
	\label{fig:MD_NN_struct}
\end{figure*}
\subsubsection{Downlink Training}
In the DL training, the objective of the UE is to find a receive beamformer ${\mathbf{S}_t}$ to maximize \eqref{eq:P-MD}.
Considering the difficulty of optimizing ${\mathbf{S}_t}$ directly, a successive solution is proposed by decomposing the original problem into multiple sub-problems, each of which can be seen as a single-beam training problem.
Specifically, ${\mathbf{S}_t}$ can be expressed as $\mathbf{S}_t=\left[ \mathbf{s}_{1,t},...,\mathbf{s}_{N_{\mathrm{s}},t} \right] $.
When $\mathbf{s}_{i,t}$ is being trained, we set the rest of the beams as zero vectors.
Correspondingly, we allocation all power to the $i$-th beam at the BS via $\left[ \mathbf{\Lambda }_{\mathrm{B},t} \right] _{i,i}=\sqrt{P_{\mathrm{B}}}$.
Therefore, when the $i$-th beam is being trained, the utility function in DL is given by 
\begin{align}
	U_{\mathrm{M}}^{\mathrm{DL}}\left( \mathbf{s}_{i,t} \right)
	&=\left| \mathbf{s}_{i,t}^{H}\mathbf{Hp}_{i,t}\mathbf{p}_{i,t}^{H}\mathbf{H}^H\mathbf{s}_{i,t}^{} \right| \notag \\
	&=\left| \mathbf{s}_{i,t}^{H}\mathbf{Hp}_{i,t} \right|^2 \notag \\
	&\approx \left| \mathbf{s}_{i,t}^{H}\mathbf{y}_{t}^{\mathrm{DL}} \right|^2.
\end{align} 
However, such an idea suffers from the fact that beams obtained via this method are not necessarily orthogonal to each other, thus leading to severe inter-beam interference.
To this end, we adopt a Gram-Schmidt method to cancel out this interference by training one beam in the orthogonal space to all the previous beams that have been trained.
Specifically, utility function $U_{\mathrm{M}}^{\mathrm{DL}}\left( \mathbf{S}_t \right)$ can be written as
\begin{align}
	\tilde{U}_{\mathrm{M}}^{\mathrm{DL}}\left( \mathbf{s}_{i,t} \right) 
	 =\left| \mathbf{s}_{i,t}^{H}\left( \mathbf{I}_M-\sum\nolimits_{p=1}^{i-1}{\mathbf{s}_{p,t}^{}\mathbf{s}_{p,t}^{H}} \right) \mathbf{y}_{t}^{\mathrm{DL}} \right|^2.
\end{align}
By following such a step, the DL multi-beam training problem can be formulated as
\begin{subequations}
	\begin{align} 
		&\max_{\mathbf{s}_{i,t}}~\tilde{U}_{\mathrm{M}}^{\mathrm{DL}}\left( \mathbf{s}_{i,t} \right)  \label{eq:MD-Obj}\\ 
		&~{\rm{s.t.}}~ \eqref{eq:P-c1}. \notag 
	\end{align}
\end{subequations}
Since for a given round $t$, only one column of $\mathbf{S}_{t}$, i.e., $\mathbf{s}_{i,t}$, is trained.
Thus, the solution of problem \eqref{eq:MD-Obj} is similar to problem \eqref{eq:SD-Obj-DL} in the single-beam case, except for two differences.
Firstly, we set a threshold denoted by $\epsilon_{\rm toler}$ to determine whether a given beam is trained sufficiently. 
Once $\epsilon_t = (\tilde{U}_{\mathrm{M}}^{\mathrm{DL}}\left( \mathbf{s}_{i,t} \right) - \tilde{U}_{\mathrm{M}}^{\mathrm{DL}}\left( \mathbf{s}_{i,t-1} \right)) / \tilde{U}_{\mathrm{M}}^{\mathrm{DL}}\left( \mathbf{s}_{i,t} \right)  < \epsilon_{\rm toler}$, it means the $i$-th beam has been trained sufficiently and the successive beam will be trained in the next round.
Secondly, we introduce a decay factor to the learning rate denoted by $\alpha$.
After one beam is trained, the learning rate is lower by $w_{\rm U} = \alpha w_{\rm U}$ since there will be a smaller space for searching.
To achieve synchronization, there is a feedback link from the UE to the BS.
When the current beam is trained sufficiently, the UE will inform the BS to start the training process of the next beam\footnote{
	According to the 5G NR beam management procedure \cite{heng2023grid}, the receiver must report the beam measurements on the transmitted beamformed reference signals to the transmitter, thus necessitating a feedback link. 
	In practice, this link can be realized using dedicated signaling channels, e.g., robust lower-frequency bands. More importantly, the feedback is only necessary when a beam is trained sufficiently, resulting in limited
	and periodical feedback requirements.
	Therefore, the bandwidth for supporting this dedicated feedback link is affordable.
}.

\subsubsection{Uplink Training}
\begin{figure}
	\centering
	\includegraphics[width=0.7\linewidth]{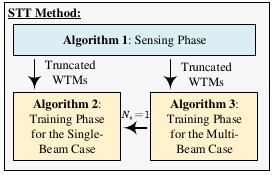}
	\caption{An illustration of the relationships between algorithms.} \label{fig:relationship_between_algorithms}
\end{figure}

In the UL training, the objective of the BS is to find a transmit beamformer ${\mathbf{P}_t}$ to maximize \eqref{eq:P-MD}.
Similar to DL training, we decompose ${\mathbf{P}_t}$ to $\mathbf{P}_t=\left[ \mathbf{p}_{1,t},...,\mathbf{p}_{N_{\mathrm{s}},t} \right] $.
When $\mathbf{p}_{i,t}$ is being trained, we first set the columns of $\mathbf{P}_t$ as zero vectors except for the $i$-th column.
Then, we allocate all transmit power at the UE to the $i$-th beam via $\left[ \mathbf{\Lambda }_{\mathrm{U},t} \right] _{i,i}=\sqrt{P_{\mathrm{U}}}$, while leaving the rest entries as zero.
When the $i$-th beam is being trained, the utility function in UL is given by 
\begin{align}
	U_{\mathrm{M}}^{\mathrm{UL}}\left( \mathbf{p}_{i,t} \right)&=\left| \mathbf{p}_{i,t}^{H}\mathbf{H}^H\mathbf{s}_{i,t}^{}\mathbf{s}_{i,t}^{H}\mathbf{H}^H\mathbf{p}_{i,t}^{} \right| \notag \\
	&=\left| \mathbf{p}_{i,t}^{T}\mathbf{H}^T\mathbf{s}_{i,t}^{*}\mathbf{s}_{i,t}^{T}\mathbf{H}^*\mathbf{p}_{i,t}^{*} \right| \notag \\
	&=\left| \mathbf{p}_{i,t}^{T}\mathbf{H}^T\mathbf{s}_{i,t}^{*} \right|^2 \notag \\
	&\approx \left| \mathbf{p}_{i,t}^{T}\mathbf{y}_{t}^{\mathrm{DL}} \right|^2.
\end{align} 
To guarantee the columns in $\mathbf{P}_t$ are orthogonal to each other, when the $i$-th beam is being trained, we can reformulate the UL objective function $U_{\mathrm{M}}^{\mathrm{UL}}\left( \mathbf{P}_t \right)$ using the Gram-Schmidt method, i.e., 
\begin{align}
	\tilde{U}_{\mathrm{M}}^{\mathrm{UL}}\left( \mathbf{p}_{i,t} \right) =\left| \mathbf{p}_{i,t}^{T}\left( \mathbf{I}_N-\sum\nolimits_{p=1}^{i-1}{\mathbf{p}_{p,t}^{}\mathbf{p}_{p,t}^{H}} \right) \mathbf{y}_{t}^{\mathrm{UL}} \right|^2.
\end{align}
By following such an idea, the UL multi-beam training problem can be formulated as
\begin{subequations}
	\begin{align} 
		&\max_{\mathbf{p}_{i,t}}~\tilde{U}_{\mathrm{M}}^{\mathrm{UL}}\left( \mathbf{p}_{i,t} \right)  \label{eq:MD-Obj-UL}\\ 
		&~{\rm{s.t.}}~ \eqref{eq:P-c2}. \notag
	\end{align}
\end{subequations}
The solution to problem \eqref{eq:MD-Obj-UL} is similar to that of \eqref{eq:SD-Obj-UL}.
In addition, after one beam is trained sufficiently, the learning rate is lowered by $w_{\rm B} = \alpha w_{\rm B}$ at the BS.
Finally, the method is shown in Fig. \ref{fig:MD_NN_struct} and summarized in {\textbf{Algorithm 3}}. \footnote{Similar to the single-beam STT, the multi-beam STT trains NNs in an online fashion as well.}
The relationship between the proposed three algorithms is illustrated by Fig. \ref{fig:relationship_between_algorithms}. 

\begin{algorithm}[ht] 
	\caption{STT: Training Phase for the Multi-Beam Case}
	\label{alg:model_based_multi_beam_alignment}
	\begin{algorithmic}[1] \small
		\STATE{\textbf{Initialization}: obtain $\mathbf{\Phi }_{\mathrm{B}}^{\left( \mathrm{e} \right)}$ and $\mathbf{\Phi }_{\mathrm{B}}^{\left( \mathrm{e} \right)}$ via \textbf{Algorithm 1}; initialize $\epsilon_{\rm toler}$, $\epsilon_0 = 0$, and $T_{\rm a}$; initialize $\mathbf{R}_{\rm U} = \mathbf{I}_M$ and $\mathbf{R}_{\rm B} = \mathbf{I}_N$, $\mathbf{P} = \mathbf{0}_{N \times N_{\rm s}}$ and $\mathbf{S} = \mathbf{0}_{M \times N_{\rm s}}$; initialize $i=1$ and $\alpha=0.99$; set learning rates $w_{\rm U}$ and $w_{\rm U}$}
		\FOR{$t = 0, 1, 2, ..., T_{\rm a}$}
		\STATE{\textbf{UE}: \makecell[l]{a)~obtain the receive beamformer  by $\mathbf{s}_{t}=g\left( \mathbf{R}_{\rm U} \mathbf{y}_{t}^{\mathrm{U}} ;\boldsymbol{\vartheta }_t \right)$; \\ b)~obtain $\boldsymbol{\vartheta }_{t+1}$ by updating  $\boldsymbol{\vartheta }_{t}$ using \eqref{loss-ue}; \\ c)~transmit $c_{\text{U}}$ using $\mathbf{s}_{t}^{*}$}}
		\STATE{\textbf{BS}: \makecell[l]{ a)~obtain the transmit beamformer by $\mathbf{p}_{t}=f\left( \mathbf{R}_{\rm B} \mathbf{y}_{t}^{\text{B}} ;\boldsymbol{\theta }_t \right)$; \\ b)~obtain $\boldsymbol{\theta }_{t+1}$ by updating  $\boldsymbol{\theta }_{t}$ using \eqref{eq:wei_up_UL}; \\ c)~transmit $c_{\text{B}}$ using $\mathbf{p}_{t}$.}}
		\STATE{Update $\left[ \mathbf{P} \right] _{:,i}=\mathbf{p}_t$ and $\left[ \mathbf{S} \right] _{:,i}=\mathbf{s}_t$ and calculate $\epsilon_t$.}
		\IF{$\epsilon_t < \epsilon_{\rm toler}$}
		\STATE Start to train the next beam pair by $i \leftarrow i+1$.
		\ENDIF
		\STATE{$\mathbf{R}_{\rm U} \leftarrow \mathbf{R}_{\rm U} - \mathbf{s}_t\mathbf{s}_{t}^{H}$ and $\mathbf{R}_{\rm B} \leftarrow \mathbf{R}_{\rm B}-\mathbf{p}_t\mathbf{p}_{t}^{H}$}
		\STATE Learning rates at the BS and UE decay by $\alpha$.
		\ENDFOR
	\end{algorithmic}
\end{algorithm}

\subsection{Stability, Required Information, and Cost Analysis} \label{sect:computaional_complexity_multi_beam}
Similar to the single-beam case, the stability relies on the received SNRs at the transceivers.
In addition, the truncated WTMs can be obtained locally for the BS and the UE.
However, in contrast to the single-beam case, the multi-beam case requires periodic information exchange.
Such a process is critical in informing the BS when a beam has been sufficiently trained.
In practice, this link can be realized using dedicated low-frequency channels.
Given the computational complexity for calculating $\epsilon_t$ is $\mathcal{O}(1)$, the computational complexity for the multi-beam STT is the same as that of the single-beam case.
However, since beams are trained successively, a larger $T_{\rm a}$ will lead to higher complexity.

\section{Numerical Results} \label{sec:result}
In this section, the performance of the proposed STT scheme is evaluated.
All the simulation results are obtained after 100 Monte Carlo simulations.
The physical-layer parameters are listed in Tab. \ref{tab:phy}.
\begin{table}[!t]
	\caption{Simulation parameters.}
	\label{tab:phy}
	\footnotesize
		\centering
		\begin{tabular}{|c|c|}
			\hline
			Transmit power at transceivers $P_{\rm B}$ and $P_{\rm U}$  & $20$ dBm\\ \hline
			Noise power spectrum density & $ -174$ dBm/Hz\\ \hline
			System bandwidth & $100$ MHz \\ \hline
			Number of antennas at transceivers $N$ and $M$   & $255$ \\ \hline
			Carrier frequency $f$ & $28$ GHz \\ \hline 
			Number of NLoS paths $L$ & $3$\\ \hline  
			Scattering loss $\alpha_l$ & $-15$ dB \\ \hline
			Transmit and receive antenna gains $G_{\rm t}$ and $G_{\rm r}$ & $15$ dB, $5$ dB \\ 
			 \hline 
		\end{tabular}
\end{table}
The scatterers are distributed uniformly between the transceivers.
For the learning parameters, the architectures of NNs at the BS and the UE are given by $N \times 128 \times 64 \times |\mathcal{G} _{\mathbf{k}}^{\left( \mathrm{e} \right)} |$ and $M \times 128 \times 64 \times |\mathcal{G} _{\boldsymbol{\kappa}}^{\left( \mathrm{e} \right)} |$, respectively.
For activation functions, linear function is used for the last layer of NNs at the BS and the UE, and ReLU function is used for the rest. 
The learning rates are set to $w_{\text{B}} = w_{\text{U}} = 0.005$, and Adam is used as the optimizer for the modules. 
The evolving rule for the decay factor $\alpha$ is given by $\alpha_{i+1} = \min \{0.001, 0.99\alpha_{i} \}$. 

\subsection{Performance of Sensing Phase}
In this sub-section, we first visualize the results obtained in the sensing phase with $T_{\rm s}=10$ and $d_{\rm BU}=15~m$.
\begin{figure} 
	\centering
	\subfloat[$|\tilde{\mathbf{H}}_{\mathrm{e}}|,~\Gamma_{\bf w}=0.1{w}_{\max}^{\rm DL/UL}$]{
		\includegraphics[width=0.47\linewidth]{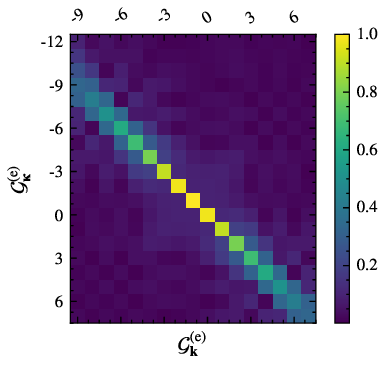}} 
	\subfloat[\textbf{$|\tilde{\mathbf{H}}_{\mathrm{e}}|,~\Gamma_{\bf w}=0.5{w}_{\max}^{\rm DL/UL}$}]{
		\includegraphics[width=0.5\linewidth]{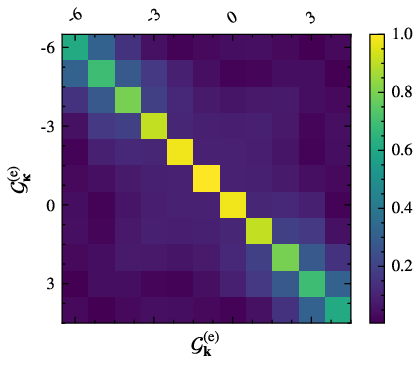}} \\
	\subfloat[\textbf{$|\tilde{\mathbf{H}}_{\mathrm{e}}|,~\Gamma_{\bf w}=0.9{w}_{\max}^{\rm DL/UL}$}]{
		\includegraphics[width=0.5\linewidth]{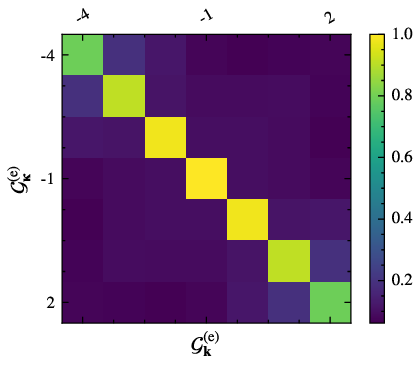}}
	\caption{The normalized wavenumber-domain channel representations under different $\Gamma_{\bf w}$.} 
	\label{fig:waveform_domain_truncate} 
\end{figure}
\begin{figure}[t]
	\centering
	\includegraphics[width=0.4\textwidth]{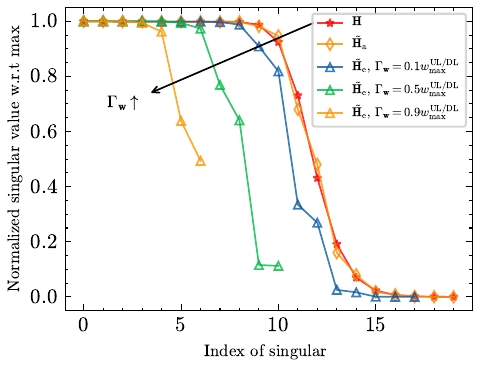}
	\caption{Normalized singular value (w.r.t the maximum value) versus the index of singular values, under different detection threshold $\Gamma_{\bf w}$.}	\label{fig:sing_vs_index_diff_thresholds}
\end{figure}

\begin{figure}[ht]
	\centering
	\includegraphics[width=0.4\textwidth]{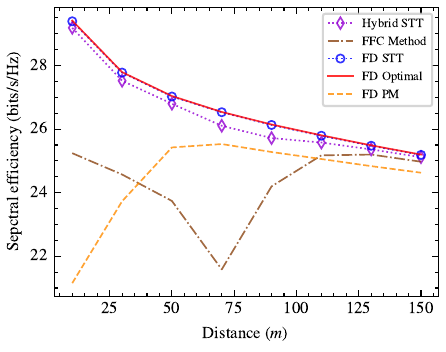}
	\caption{Throughput versus distance in meters with $T_{\rm a}=125$.}
	\label{fig:throughput_distance}
\end{figure} 
In Fig. \ref{fig:waveform_domain_truncate}, the wavenumber-domain channel representations are plotted using WTMs.
In this figure, ${w}_{\max}^{\rm DL}$ and ${w}_{\max}^{\rm UL}$ denote the most significant entries of $\hat{\bf w}^{\rm DL}$ and $\hat{\bf w}^{\rm UL}$, respectively.
As shown by Fig. \ref{fig:H_a}, the channel representation in the wavenumber domain, i.e., $\tilde{\mathbf{H}}_{\mathrm{a}}$, is sparse and diagonal.
Then, using truncated WTMs, i.e., $\mathbf{\Phi }_{\mathrm{U}}^{\left( \mathrm{e} \right)}$ and $\mathbf{\Phi }_{\mathrm{B}}^{\left( \mathrm{e} \right)}$, the LoS sub-space can be extracted, which can reduce the channel dimension.
By tuning $\Gamma_{\bf w}$ larger, we can observe that the dimension of $\tilde{\mathbf{H}}_{\mathrm{e}}$ decreases correspondingly.
The reason is that a smaller sub-space composed of more significant values is extracted by a larger $\Gamma_{\bf w}$. 

In Fig. \ref{fig:sing_vs_index_diff_thresholds}, we verify the effectiveness of the wavenumber-domain analysis by using singular value decomposition (SVD).  
For simplicity of analysis, the singular values are normalized according to the most significant entries.
It can be seen from Fig. \ref{fig:sing_vs_index_diff_thresholds} that, unlike the rank-$1$ far-field LoS channel, the near-field channel has a higher rank even in a scatter-sparse environment, i.e., $L=3$.
Illustratively, $\mathbf{H}$ and $\tilde{\mathbf{H}}_{\mathrm{a}}$ have near-identical normalized singular values since they are semi-unitary equivalent, indicating that there is no information loss when the channel information is transformed to the wavenumber domain.
By increasing $\Gamma_{\bf w}$, we can observe from Fig. \ref{fig:sing_vs_index_diff_thresholds} that fewer singular values are included in a smaller sub-space of the wavenumber domain.   
Thus, there is a tradeoff between the dimension and the number of available DoFs. 
Additionally, by choosing a proper $\Gamma_{\bf w}$, the top singular values can be preserved in the truncated wavenumber domain.   

\subsection{Performance of Training Phase}
In this sub-section, we investigate the performance of the proposed hybrid STT scheme under the single-beam and multi-beam cases.
The following four benchmarks are considered in our simulation:
\begin{itemize}
	\item {\textbf{Far-field Codebook (FFC) Method} \cite{he2015suboptimal}:} In this benchmark, the angular space is traversed by binary-tree-based beam searching in a coarse-to-fine manner, while the unit modulus constraint is considered.
	Since such a method cannot extend to the multi-beam case, we use this as a benchmark for the single-beam case.
	
	\item {\textbf{Fully-digital (FD) Opt.}}: By assuming the perfect channel information $\bf H$ is known, this method is obtained by SVD, i.e., $\mathbf{P}=\mathbf{U} \in \mathbb{C}^{N \times N_{\rm s}}$ and $\mathbf{S}=\mathbf{V} \in \mathbb{C}^{M \times N_{\rm s}}$, where $\mathbf{U}$ and $\mathbf{V}$ denote $N_{\rm s}$ most significant left and right singular vectors of $\bf H$. 
	This method is realized using the FD beamforming technique.
	
	\item {\textbf{FD STT}}: In this method, we relax the unit modulus constraint by adopting the FD architecture. 
	
	\item {\textbf{FD Power Method (PM)}  \cite{Dahl2004blind}}: The method adopts ping-pong pilots to actively estimate the top singular vectors of a MIMO channel in an iterative manner.
	This method is implemented using the FD beamforming technique. 
\end{itemize}

\begin{figure}[t!]
	\centering
	\includegraphics[width=0.4\textwidth]{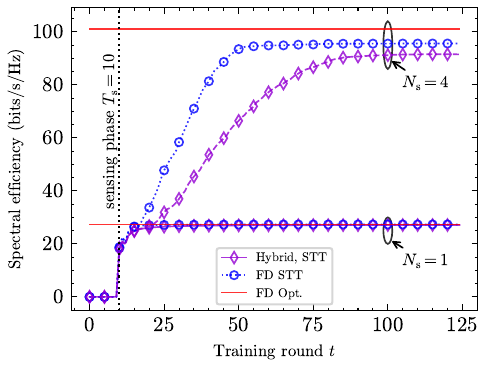}
	\caption{Spectral efficiency versus $t$ under $N_{\rm s}=1~{\rm and}~4$.} \label{fig:SE_vs_Ta_diff_Ns}
\end{figure}
\begin{figure}[t!]
	\centering
	\includegraphics[width=0.4\textwidth]{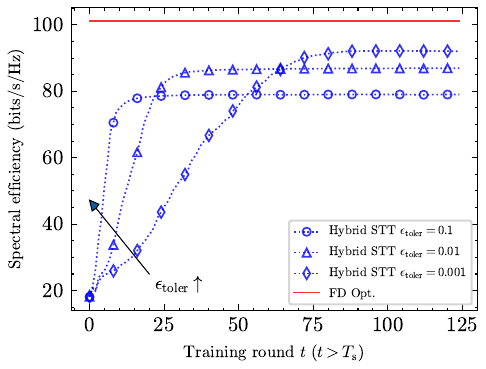}
	\caption{Spectral efficiency versus $t~(t>T_{\rm s})$ under different $\epsilon_{\rm toler}$, with fixed $N_{\rm s}=4$.}\label{fig:se_vs_T_a_diff_toler}
\end{figure}
Fig. \ref{fig:throughput_distance} illustrates the performance of the proposed scheme against other benchmarks in the single-beam case.
In this figure, with a fixed training round $T_{\rm a}$, the proposed hybrid STT scheme can achieve a near-optimal performance with a gap incurred by the unit modulus constraint.
Therefore, by relaxing the unit modulus constraint, the FD STT scheme can realize the near-optimal SE.
For the conventional method, the FD PM method cannot provide an acceptable SE when the transceivers are close.
The reason is that the spherical wavefront in the near field can provide more DoFs, making the conventional schemes unable to find the optimal beam pair quickly.
Lastly, the FFC scheme fails to provide a decent SE in the near field since it ignores the distance dependence of near-field channels.
On the contrary, the proposed scheme is applicable to both the near-field and far-field scenarios.

In Fig. \ref{fig:SE_vs_Ta_diff_Ns}, the beam training process is shown for the single-beam ($N_{\rm s}=1$) and multi-beam ($N_{\rm s}=4$) cases.
It is noted that as a result of the STT scheme, the training process begins after sensing, i.e., $T_{\rm s}=10$.  
This figure illustrates that the proposed hybrid STT scheme can achieve near-optimal results in the single-beam case while having a larger gap to the optimal in the multi-beam case.
The reason is two-fold. 1)~\emph{Accumulation of errors}: in the multi-beam case, each beam keeps being trained until the error falls below the tolerable threshold, i.e., $\epsilon_t < \epsilon_{\rm toler}$.
Therefore, the error existing for each beam will accumulate and harm the achieved SE.
2)~\emph{Correlation among beams}: according to line 9 \textbf{Algorithm} \ref{alg:model_based_multi_beam_alignment}, a beam is trained in the orthogonal space spanned by the former beams to guarantee the orthogonality among beams.
However, due to noisy observations of the pilots at transceivers, orthogonality among beams can be interfered, thus causing a degradation of the achieved SE.
On the contrary, in the single-beam case, only one beam pair needs to be trained so that the aforementioned problems can be avoided, resulting in a smaller gap to the optimal results.
Moreover, compared to the single-beam case, the multi-beam case requires a longer time to carry out beam training.
Compared to the FD STT scheme, the proposed hybrid STT scheme takes more time to converge.
This is because without the unit modulus constraint, the FD STT scheme can train the beam with higher flexibility, thus accelerating the training process.
Lastly, the unit modulus constraint is also the origin of the gap between the hybrid and FD STT schemes. 

In Fig. \ref{fig:se_vs_T_a_diff_toler}, we vary the tolerable threshold. i.e., $\epsilon_{\rm toler}$, to further study \emph{accumulation of errors}. 
Illustratively, as $\epsilon_{\rm toler}$ climbs from $0.001$ to $0.1$, there will be a larger gap between the proposed hybrid STT scheme and the optimal one.
This is because more errors will accumulate more for a larger threshold, thus deteriorating the SE performance.
These results are consistent with our analysis of Fig. \ref{fig:SE_vs_Ta_diff_Ns}.
Furthermore, with a smaller threshold, the proposed STT scheme needs more time to converge since individual beams are trained more finely.

\begin{figure}[t!]
	\centering
	\subfloat[BS beam index 1]{
		\includegraphics[width=0.5\linewidth]{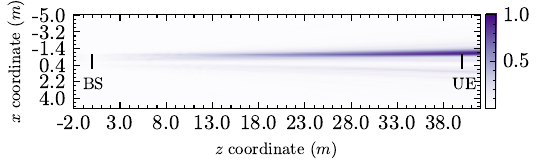}}
	\subfloat[UE beam index 1]{
		\includegraphics[width=0.5\linewidth]{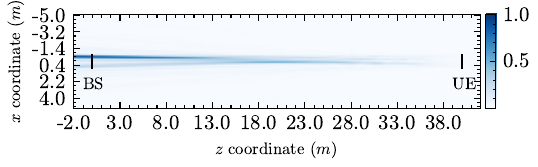}}
	\\
	\subfloat[BS beam index 2]{
		\includegraphics[width=0.5\linewidth]{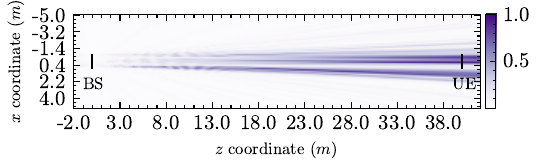}}
		\subfloat[UE beam index 2]{
		\includegraphics[width=0.5\linewidth]{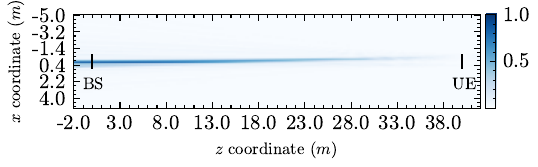}}
	\\
	\subfloat[BS beam index 3]{
		\includegraphics[width=0.5\linewidth]{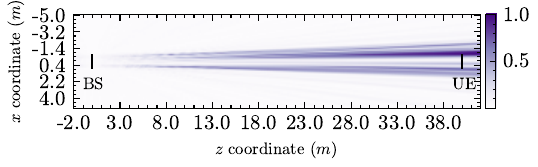}}
		\subfloat[UE beam index 3]{
		\includegraphics[width=0.5\linewidth]{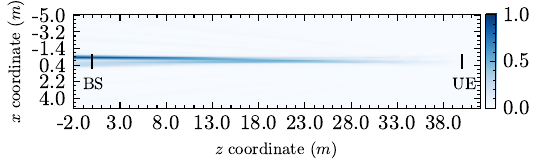}}
	\\
	\subfloat[BS beam index 4]{
		\includegraphics[width=0.5\linewidth]{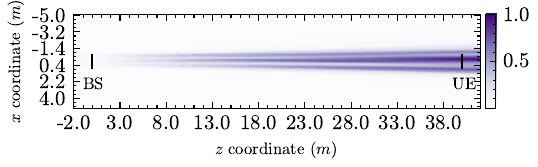}}
	\subfloat[UE beam index 4]{
		\includegraphics[width=0.5\linewidth]{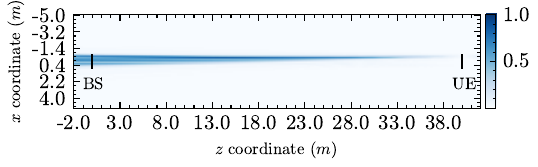}}
	\caption{The beamfocusing performance of the BS and the UE when $N_{\rm s}=4$. Beam gains are normalized according to the maximum value.} \label{fig:beamfocosing_performance}
\end{figure}
In Fig. \ref{fig:beamfocosing_performance}, we visualize the beam training results for the multi-beam case.
Specifically, we adopt the columns in $\mathbf{S}$ and $\mathbf{P}$ to calculate their gains with respect to the array response vectors of a given position ${\bf v} \in \mathbb{R}^{3 \times 1}$, which are defined as
$\mathbf{a}_{\mathrm{B}}\left( \mathbf{v} \right) =\left[ e^{-jk_0\left\| \mathbf{v}-\mathbf{x}_{-\tilde{N}} \right\|},...,e^{-jk_0\left\| \mathbf{v}-\mathbf{x}_{\tilde{N}} \right\|} \right] ^T$ and 
$\mathbf{a}_{\mathrm{U}}\left( \mathbf{v} \right) =\left[ e^{-jk_0\left\| \mathbf{v}-\mathbf{r}_{-\tilde{M}} \right\|},...,e^{-jk_0\left\| \mathbf{v}-\mathbf{r}_{\tilde{M}} \right\|} \right] ^T
$ for the BS and the UE, respectively.
As shown by the figures, for the BS, the beams are focused at the location of the UE, while for the UE, the beams are focused at the location of the BS.
It can be observed from the figure that the beams are not focusing on a spot.
The reason is that the physical sizes of antenna arrays are not negligible in the near field.
Therefore, since MIMO is considered in our work, the beams should focus on the entry antenna arrays instead of a spot. 
Additionally, we can also see some mis-focusing beams, which are the origin of the performance loss.

\begin{figure} 
	\centering
	\includegraphics[width=0.41\textwidth]{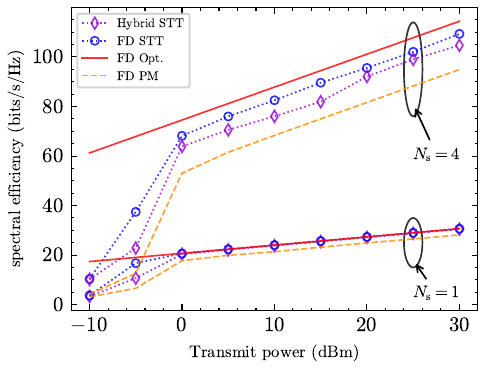}
	\caption{SE versus transmit power $P$ with $d_{\rm BU}=40~m$, $N_{\rm s}=4$, $T_{\rm s}=10$, and $T_{\rm a}=125$.} \label{fig:se_vs_p}
\end{figure}
In Fig. \ref{fig:se_vs_p}, we present the achieved SE against the transmit power for $d_{\rm BU}=40~m$, $T_{\rm s}=10$, and $T_{\rm a}=125$.
When the transmit power is larger than $ 0 $ dBm, the hybrid STT scheme can achieve a near-optimal performance.
Like the above figures, the unit modulus constraint incurs the gap between the proposed hybrid STT and the FD STT.
However, the gap between the proposed and optimal schemes is more significant in the low power region, i.e., $P<0~{\rm dBm}$.
The reason is that the noisy received pilots can mislead the learning process of the transceivers. 
It is vital to notice that the proposed hybrid STT scheme consistently outperforms the conventional PM method for any transmit power, validating the critical role that NNs play.
Additionally, it is interesting to observe that within the transmit power ranging from $P=-10~{\rm dBm}$ to $P=-0~{\rm dBm}$, the gap between the proposed hybrid STT and the FD STT is larger.
It shows that the FD STT scheme is more robust to the noise without the unit modulus constraint.

\begin{figure}[t]
	\centering
	\includegraphics[width=0.4\textwidth]{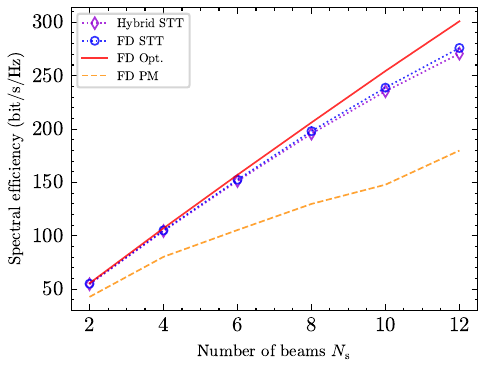}
	\caption{Spectral efficiency versus $N_{\rm s}$ with $d_{\rm BU}=15~m$, $T_{\rm s}=10$, and $T_{\rm a}=250$.} \label{fig:SE_vs_Ns}
\end{figure}

\begin{figure}[t]
	\centering
	\includegraphics[width=0.4\textwidth]{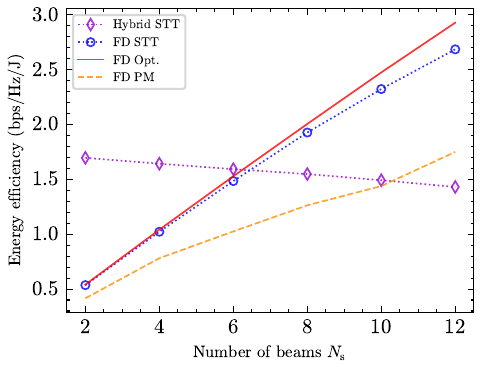} 
	\caption{Energy efficiency versus $N_{\rm s}$ with $d_{\rm BU}=15~m$, $T_{\rm s}=10$, and $T_{\rm a}=250$.} \label{fig:EE_vs_Ns}
\end{figure}

In Fig. \ref{fig:SE_vs_Ns}, we investigate the achieved SE by varying the number of beams $N_{\rm s}$. 
In this setup, the effective DoF (EDoF) of the MIMO channel is $12.68$, which represents the number of independent channels that a MIMO channel can be decomposed to. 
The value of EDoF can be calculated by $\mathrm{EDoF}\approx \mathrm{tr}( \mathbf{C}) ^2/\mathrm{tr}( \mathbf{C}^2 )$, where $ \mathbf{C}=\mathbf{H}^H\mathbf{H}$ \cite{xie2023performance}.
To guarantee that all the methods converge, we extend $T_{\rm a}$ to $250$ and run the simulation under $N_{\rm s}=2,4,...,12$.
This figure demonstrated that as $N_{\rm s}$ increases, the achieved SE will increase correspondingly, with an increasingly large gap to the optimal.
This can be attributed to \emph{accumulation of errors} and \emph{correlation among beams} mentioned before.
In practice, learned from the results in Fig. \ref{fig:se_vs_T_a_diff_toler}, this gap can be narrowed by introducing stricter $\epsilon_{\rm toler}$, at the cost of training overheads.
These issues are more prominent when $N_{\rm s}$ is large, since the EDoF is used up.
In contrast, for the conventional PM method, even though the training phase is extended, it still cannot provide decent performance.
  
In Fig. \ref{fig:EE_vs_Ns}, we investigate the achieved EE by varying the number of beams $N_{\rm s}$ under the same settings of Fig. \ref{fig:SE_vs_Ns}.
Firstly, we analyze the energy consumption of the multi-beam STT scheme.
Since the ping-pong pilots are utilized for training, it is reasonable to quantify the power consumption of one round.
Therefore, the total power consumption can be expressed as $P_{\rm sum} = P_{\rm B} + P_{\rm U} + P_{\rm RF} (N_{\rm RF, B} + N_{\rm RF, U}) + 2P_{\rm BB}+ P_{\rm PS} (N_{\rm PS, B} + N_{\rm PS, U}) 
$, where $P_{\rm RF}$, $P_{\rm PS}$, and $P_{\rm BB}$ denote the power consumption of an RF chain, that of a PS, and that of baseband processing, respectively.
$N_{\rm RF, B}$ and $N_{\rm PS, B}=N\times N_{\rm RF, B}$ denote the number of RF chains and that of PS at the BS, respectively.
Lastly, $N_{\rm RF, U}$ and $N_{\rm PS, U}=M\times N_{\rm RF, U}$ denote the number of RF chains and that of PS at the UE, respectively.
Here, we set $P_{\rm RF}=200$ mW, $P_{\rm PS}=30$ mW, and $P_{\rm BB}=300$ mW.
The energy efficiency (EE) for the multi-beam case is calculated by $E\left( \mathbf{S},\mathbf{P} \right) =R\left( \mathbf{S},\mathbf{P} \right) /P_{\mathrm{sum}}$.
For the FD configurations, we can see that their EE increases with $N_{\rm s}$ in a linear fashion.
This is because, for the FD configuration, we have $N_{\rm RF,B}=N$, $N_{\rm RF,U}=M$, and $N_{\rm PS}=0$, which make $P_{\rm sum}$ a constant value.
Hence, the behavior of EE curves is in line with that of its SE curve.
In contrast, the EE of hybrid STT decreases with $N_{\rm s}$, which can be explained by the following.
Initially, since PSs consume less energy than the RF chains, the hybrid STT is more energy efficient than the FD counterpart.
However, when $N_{\rm s}$ continues to increase, the energy consumption of the increased PSs has a dominant impact, thus leading to a decrease in EE.
Therefore, our method can strike a good balance between throughput and energy cost when $N_{\rm s}$ is small.

\section{Conclusion} \label{sec:conclusion}
In this paper, we proposed an STT scheme to realize beam training for both the single- and multi-beam cases for near-field MIMO systems.
To be specific, during the sensing phase, the truncated WTMs are obtained locally by sensing, with which a low-dimensional subspace in the wavenumber domain can be extracted.
Then, in the subsequent beam training phase, the NN modules at the transceivers were updated based on incoming ping-pong pilots and trained incrementally with online data points.	
Simulation results validated that the proposed STT scheme
enables fast and low-dimensional beam training for both
cases while achieving performance close to the
optimal method, which relies on perfect CSI.

\bibliographystyle{IEEEtran}
\bibliography{reference/mybib}

\begin{thebibliography}{10}
\providecommand{\url}[1]{#1}
\csname url@samestyle\endcsname
\providecommand{\newblock}{\relax}
\providecommand{\bibinfo}[2]{#2}
\providecommand{\BIBentrySTDinterwordspacing}{\spaceskip=0pt\relax}
\providecommand{\BIBentryALTinterwordstretchfactor}{4}
\providecommand{\BIBentryALTinterwordspacing}{\spaceskip=\fontdimen2\font plus
\BIBentryALTinterwordstretchfactor\fontdimen3\font minus
  \fontdimen4\font\relax}
\providecommand{\BIBforeignlanguage}[2]{{%
\expandafter\ifx\csname l@#1\endcsname\relax
\typeout{** WARNING: IEEEtran.bst: No hyphenation pattern has been}%
\typeout{** loaded for the language `#1'. Using the pattern for}%
\typeout{** the default language instead.}%
\else
\language=\csname l@#1\endcsname
\fi
#2}}
\providecommand{\BIBdecl}{\relax}
\BIBdecl

\bibitem{jiang2024active}
H.~Jiang, Z.~Wang, and Y.~Liu, ``Active-sensing-based beam alignment for near
  field {MIMO} communications,'' in \emph{Proc. IEEE Intl. Conf. Commun.
  (ICC)}, Accepted to appear, Jun. 2024.

\bibitem{jiang2021road}
W.~Jiang, B.~Han, M.~A. Habibi, and H.~D. Schotten, ``The road towards {6G}: A
  comprehensive survey,'' \emph{IEEE Open J. Commun. Soc.}, vol.~2, pp.
  334--366, 2021.

\bibitem{itu2015imt}
I.-R.~S. M.2370-0, ``{IMT} traffic estimates for the years 2020 to 2030,'' Jun.
  2015.

\bibitem{Shafie2023Therahertz}
A.~Shafie, N.~Yang, C.~Han, J.~M. Jornet, M.~Juntti, and T.~Kürner,
  ``Terahertz communications for {6G} and beyond wireless networks: Challenges,
  key advancements, and opportunities,'' \emph{IEEE Netw.}, vol.~37, no.~3, pp.
  162--169, May/Jun. 2023.

\bibitem{wang2018millimeter}
X.~Wang, L.~Kong, F.~Kong, F.~Qiu, M.~Xia, S.~Arnon, and G.~Chen, ``Millimeter
  wave communication: A comprehensive survey,'' \emph{IEEE Commun. Surveys
  Tuts.}, vol.~20, no.~3, pp. 1616--1653, Jun. 2018.

\bibitem{jiang2023active}
T.~Jiang, F.~Sohrabi, and W.~Yu, ``Active sensing for two-sided beam alignment
  and reflection design using ping-pong pilots,'' \emph{IEEE J. Sel. Areas
  Info. Theory}, vol.~4, pp. 24--39, May 2023.

\bibitem{nitsche2015steering}
T.~Nitsche, A.~B. Flores, E.~W. Knightly, and J.~Widmer, ``Steering with eyes
  closed: Mm-wave beam steering without in-band measurement,'' in \emph{Proc.
  IEEE Conf. Comput. Commun. (INFOCOM)}, Aug. 2015, pp. 2416--2424.

\bibitem{qurratulain2023machine}
M.~Qurratulain~Khan, A.~Gaber, P.~Schulz \emph{et~al.}, ``Machine learning for
  millimeter wave and terahertz beam management: A survey and open
  challenges,'' \emph{IEEE Access}, vol.~11, pp. 11\,880--11\,902, Feb. 2023.

\bibitem{heng2023grid}
Y.~Heng and J.~G. Andrews, ``Grid-free {MIMO} beam alignment through
  site-specific deep learning,'' \emph{IEEE Trans. Wireless Commun.}, Early
  Access, Jun. 2023, doi: 10.1109/TWC.2023.3283475.

\bibitem{Giordani2019tutorial}
M.~Giordani, M.~Polese, A.~Roy, D.~Castor, and M.~Zorzi, ``A tutorial on beam
  management for {3GPP} {NR} at mmwave frequencies,'' \emph{IEEE Commun.
  Surveys Tuts.}, vol.~21, no.~1, pp. 173--196, Firstquarter 2019.

\bibitem{liu2023near}
Y.~Liu, Z.~Wang, J.~Xu, C.~Ouyang, X.~Mu, and R.~Schober, ``Near-field
  communications: A tutorial review,'' \emph{IEEE Open J. Commun. Soc.},
  vol.~4, pp. 1999--2049, Aug. 2023.

\bibitem{wu2023two}
C.~Wu, C.~You, Y.~Liu, L.~Chen, and S.~Shi, ``Two-stage hierarchical beam
  training for near-field communications,'' \emph{IEEE Trans. Veh. Tech.}, pp.
  1--13, Early Access, Sept. 2023, doi: 10.1109/TVT.2023.3311868.

\bibitem{liu2024nearfield}
Y.~Liu, C.~Ouyang, Z.~Wang, J.~Xu, X.~Mu, and A.~L. Swindlehurst, ``Near-field
  communications: A comprehensive survey,'' \emph{arXiv preprint
  arXiv:2401.05900}, Jan. 2024.

\bibitem{song2017common}
J.~Song, J.~Choi, and D.~J. Love, ``Common codebook millimeter wave beam
  design: Designing beams for both sounding and communication with uniform
  planar arrays,'' \emph{IEEE Trans. Commun.}, vol.~65, no.~4, pp. 1859--1872,
  Apr. 2017.

\bibitem{zhang2020beam}
J.~Zhang, Y.~Huang, Y.~Zhou, and X.~You, ``Beam alignment and tracking for
  millimeter wave communications via bandit learning,'' \emph{IEEE Trans.
  Commun.}, vol.~68, no.~9, pp. 5519--5533, Sept. 2020.

\bibitem{wang2023improving}
S.~Wang and S.~Bi, ``Improving beam alignment accuracy in mmwave communication
  systems with auxiliary tasks,'' \emph{IEEE Sig. Process. Lett.}, vol.~30, pp.
  992--996, Jul. 2023.

\bibitem{xiao2026hierarchical}
Z.~Xiao, T.~He, P.~Xia, and X.-G. Xia, ``Hierarchical codebook design for
  beamforming training in millimeter-wave communication,'' \emph{IEEE Trans.
  Wireless Commun.}, vol.~15, no.~5, pp. 3380--3392, Jan. 2016.

\bibitem{qi2020hierarchical}
C.~Qi, K.~Chen, O.~A. Dobre \emph{et~al.}, ``Hierarchical codebook-based
  multiuser beam training for millimeter wave massive {MIMO},'' \emph{IEEE
  Trans. Wireless Commun.}, vol.~19, no.~12, pp. 8142--8152, Sept. 2020.

\bibitem{he2015suboptimal}
T.~He and Z.~Xiao, ``Suboptimal beam search algorithm and codebook design for
  millimeter-wave communications,'' \emph{Mobile Netw. Appl.}, vol.~20, pp.
  86--97, Feb. 2015.

\bibitem{li2019explore}
M.~Li, C.~Liu, S.~V. Hanly, I.~B. Collings, and P.~Whiting, ``Explore and
  eliminate: Optimized two-stage search for millimeter-wave beam alignment,''
  \emph{IEEE Trans. Wireless Commun.}, vol.~18, no.~9, pp. 4379--4393, Jun.
  2019.

\bibitem{sohrabi2022active}
F.~Sohrabi, T.~Jiang, W.~Cui, and W.~Yu, ``Active sensing for communications by
  learning,'' \emph{IEEE J. Sel. Areas Commun.}, vol.~40, no.~6, pp.
  1780--1794, Jun. 2022.

\bibitem{las2006evaluating}
F.~Las-Heras, M.~Pino, S.~Loredo, Y.~Alvarez, and T.~Sarkar, ``Evaluating
  near-field radiation patterns of commercial antennas,'' \emph{IEEE Trans.
  Antennas and Propag.}, vol.~54, no.~8, pp. 2198--2207, Aug. 2006.

\bibitem{cui2022channel}
M.~Cui and L.~Dai, ``Channel estimation for extremely large-scale {MIMO}:
  Far-field or near-field?'' \emph{IEEE Trans. Commun.}, vol.~70, no.~4, pp.
  2663--2677, Jan. 2022.

\bibitem{zhang2022fast}
Y.~Zhang, X.~Wu, and C.~You, ``Fast near-field beam training for extremely
  large-scale array,'' \emph{IEEE Wireless Commun. Lett.}, vol.~11, no.~12, pp.
  2625--2629, Oct. 2022.

\bibitem{zhang2023beam}
X.~Zhang, H.~Zhang, J.~Zhang, C.~Li, Y.~Huang, and L.~Yang, ``Codebook design
  for extremely large-scale {MIMO} systems: Near-field and far-field,''
  \emph{IEEE Trans. Commun.}, early access, Nov. 2023, doi:
  {10.1109/TCOMM.2023.3329224}.

\bibitem{cui2023near-rainbow}
M.~Cui, L.~Dai, Z.~Wang, S.~Zhou, and N.~Ge, ``Near-field rainbow: Wideband
  beam training for {XL-MIMO},'' \emph{IEEE Trans. Wireless Commun.}, vol.~22,
  no.~6, pp. 3899--3912, Jun. 2023.

\bibitem{ouyang2023nearfield}
C.~Ouyang, Y.~Liu, X.~Zhang, and L.~Hanzo, ``Near-field communications: A
  degree-of-freedom perspective,'' \emph{arXiv preprint arXiv:2308.00362}, Aug.
  2023.

\bibitem{wang2023near}
Z.~Wang, X.~Mu, and Y.~Liu, ``Near-field integrated sensing and
  communications,'' \emph{IEEE Commun. Lett.}, vol.~27, no.~8, pp. 2048--2052,
  May, 2023.

\bibitem{ROTHMAN20134}
L.~Rothman, I.~Gordon, Y.~Babikov \emph{et~al.}, ``The {HITRAN2012} molecular
  spectroscopic database,'' \emph{J. Quant. Spectrosc. Radiati. Transf.}, vol.
  130, pp. 4--50, 2013.

\bibitem{tang2023line}
A.~Tang, J.-B. Wang, Y.~Pan, W.~Zhang, Y.~Chen, H.~Yu, and R.~C. de~Lamare,
  ``Line-of-sight extra-large {MIMO} systems with angular-domain processing:
  Channel representation and transceiver architecture,'' \emph{IEEE Trans.
  Commun.}, vol.~72, no.~1, pp. 570--584, Oct. 2024.

\bibitem{pizzo2021fourier}
A.~Pizzo, L.~Sanguinetti, and T.~L. Marzetta, ``Fourier plane-wave series
  expansion for holographic {MIMO} communications,'' \emph{IEEE Trans. Wireless
  Commun.}, vol.~21, no.~9, pp. 6890--6905, Mar. 2022.

\bibitem{wei2022multi}
L.~Wei, C.~Huang, G.~C. Alexandropoulos, W.~E.~I. Sha, Z.~Zhang, M.~Debbah, and
  C.~Yuen, ``Multi-user holographic {MIMO} surfaces: Channel modeling and
  spectral efficiency analysis,'' \emph{IEEE J. Sel. Topics Signal Process.},
  vol.~16, no.~5, pp. 1112--1124, May, 2022.

\bibitem{pizzo2020holographic}
A.~Pizzo, T.~Marzetta, and L.~Sanguinetti, ``Holographic mimo communications
  under spatially-stationary scattering,'' in \emph{Proc. 54th Asilomar Conf.
  Signals, Syst. Comput.}, Nov. 2020, pp. 702--706.

\bibitem{yu2016alternating}
X.~Yu, J.-C. Shen, J.~Zhang \emph{et~al.}, ``Alternating minimization
  algorithms for hybrid precoding in millimeter wave {MIMO} systems,''
  \emph{IEEE J. Sel. Topics Signal Process.}, vol.~10, no.~3, pp. 485--500,
  Feb. 2016.

\bibitem{Dahl2004blind}
T.~Dahl, N.~Christophersen, and D.~Gesbert, ``Blind {MIMO} eigenmode
  transmission based on the algebraic power method,'' \emph{IEEE Trans. Signal
  Process.}, vol.~52, no.~9, pp. 2424--2431, Sept. 2004.

\bibitem{xie2023performance}
Z.~Xie, Y.~Liu, J.~Xu, X.~Wu, and A.~Nallanathan, ``Performance analysis for
  near-field {MIMO}: Discrete and continuous aperture antennas,'' \emph{IEEE
  Wireless Commun. Lett.}, vol.~12, no.~12, pp. 2258--2262, Dec. 2023.

\end{thebibliography}

\end{document}